# Uncovering MIT wind myths through micro-climatological CFD analysis

Alexander Kalmikov

Department of Earth, Atmospheric, and Planetary Sciences,
Massachusetts Institute of Technology, Cambridge, MA

**Abstract**

Popular campus myths of unusually strong pedestrian level winds are investigated with a Computational Fluid Dynamics (CFD) approach. The numerical simulations confirm the existence of the reported phenomena and provide a qualitative explanation of their physical mechanisms.

## 1 Introduction

Generations of MIT students remember well the notorious cold gales penetrating under their coats during the long walk back to the dorm after another tiring day working hard in the lab. The exceptionally strong winds blowing along Amherst Alley, also known as "Dorm Row," become particularly strong and gusty near MacGregor House - an undergraduate dormitory that has given its name to the familiar MIT wind phenomenon: the "MacGregor Wind Tunnel." Although the existence of this phenomenon is firmly established in MIT folklore (e.g. Balsley 1994), its physical mechanism has not been understood. Recently, in actual wind tunnel experiments Wannaphahoon (2011) measured the local acceleration of air flow around the corner of a scaled-down model of MacGregor House, but has not offered an explanation for this wind speed increase. The computational approach pursued in the present work is applied to model these high wind conditions numerically and explain their physical mechanisms through analysis of the simulated flow fields. The understanding of the physics of MacGregor high winds and their relation to the background wind environment enables prediction of the "MacGregor Wind Tunnel" occurrence.

Another well-known MIT wind myth is the story associated with the installation of the "La Grande Voile" ("The Big Sail") stabile in front of MIT's Cecil and Ida Green

Building. The myth claims that the sculpture was installed to block the strong winds blowing from Charles River across the McDermott Court and to alleviate the effect of strong wind at the base of the Green Building (Gleitzman 2006, Bourzac 2006). The wind is described to be so strong as to make it hard to open the entry doors and to rip off eyeglasses from a pedestrian's head (Crowley 2005). In fact, the occurrence of exceptionally strong winds around Green Building is well documented (Bicknell 1965, Durgin 1992). Being the tallest building in Cambridge and hosting MIT's Meteorology Department (today MIT Department of Earth, Atmospheric and Planetary Sciences) added to public interest, especially when windows fell out from top floors and a door blew in and was boarded up in the first winter after the building's construction (Burke and McCaffrey 1965). Even today in particularly windy weather, an unusually strong wind resistance can be felt when trying to open the building's swinging doors when the revolving doors are locked afterhours. So as fierce winds continue to blow occasionally through the lobby at the base of the Green Building, the questions arise - how strong were these winds before “The Big Sail“ installation, why did MIT authorities deny having planned “The Big Sail“ installation as a wind blocking solution (Gleitzman 2006, Bourzac 2006), and why is the wind so strong at the base of the Green Building?

Although myths and rumors seem to have lives of their own, the technical answer to these questions lies in the realm of Wind Engineering - the discipline that can explain the physics behind high winds in urban environments. This paper investigates the wind myths of the MIT campus with a combination of Computational Fluid Dynamics (CFD) and climatological approaches. The former allows a detailed three-dimensional analysis of wind flows and forces shaping them, revealing the physical mechanisms of local wind phenomena. The latter analyses the local statistics of larger scale atmospheric flows and weather systems. The combination of the approaches integrates the high resolution local flow details with the statistics of background flow conditions, enabling a local micro-climatological analysis. In the following chapters we overview our methodology, present results of CFD simulations and discuss the physics behind MIT wind myths.

This study was carried out as part of a larger effort to assess and map the wind resource on the MIT campus for possible installation of small wind turbines (Kalmikov et al. 2010). The author was a member of a student team consulting MIT Facilities on wind resource assessment and planning towards installation of a wind turbine on the MIT campus. A dozen anemometers have been installed around the campus to measure wind around the complex geometry of its buildings. The CFD study was initiated to extend the measured data coverage and allow a physical insight into the observed wind patterns, with the goal of optimizing turbine siting. The current paper is a side product of this effort and an exciting opportunity to resolve the longstanding controversies about MIT wind myths.

# 2 Methodology

Wind flow over the MIT campus was analyzed with a realistic model of campus geometry. A three dimensional model of the campus was generated based on the GIS* data from City of Cambridge GIS. The data was downloaded from the MIT Geodata Repository – the GeoWeb and analyzed with ArcGIS software suit (Figure 2.1). The data sources for the geometry of building roofs, terrain topography and Charles River outlines are, respectively: Cambridge building structures (Buildings, 2009); Digital Terrain Model of City of Cambridge (DEM, 2003); hydrographic features (Hydro, 2007). The data was manually edited to correct inconsistencies and the detailed terrain topography was simplified to generate an idealized topography of the river banks (Figure 2.2).

Three dimensional wind flow fields were modeled with a Virtual Wind Tunnel methodology - i.e. computer simulations of the experimental practice of placing a scaled down model of building geometry in a wind tunnel (Bicknell 1965, Wannaphahoon 2011). In wind tunnel experiments the wind blows from a predetermined inlet direction and exits the experimental area from the opposite side, while the scaled model of the buildings is rotated with different angles to the inflow direction. Here the CFD simulations were performed with the Meteodyn model UrbaWind (Fahssis et al. 2010), developed for computing wind energy and pedestrian wind comfort in urban environment (Caniot et al. 2011). The turbulent air motion is resolved with Reynolds-Averaged Navier–Stokes equations (RANS); the sub-grid turbulent fluxes are modeled with k-L parameterization. Only a steady flow solution is obtained, air thermodynamics neglected, the flow assumed incompressible with constant density. The computational mesh was generated automatically in UrbaWind for each computed direction aligned with the wind flow. The grid is unstructured Cartesian with automatic refinement near the ground and the obstacles (see Figure 2.3). A symmetry condition is applied at the lateral boundaries, a reflecting condition at the upper boundary and a homogeneous pressure condition at the outlet. The effect of porous obstacles is modeled by introducing a momentum sink term in the cells lying inside the obstacle; its strength given by a volumetric friction coefficient

* GIS - Geographic Information System, is a modern cartography technology and digital format.

$C_D$. Details of this parameterization and the numerics of the CFD model are summarized in Kalmikov et al. (2010). Further description and experimental validation of the code with Architectural Institute of Japan (AIJ) data (Tominaga et al. 2008) is given in Fahssis et al. (2010).

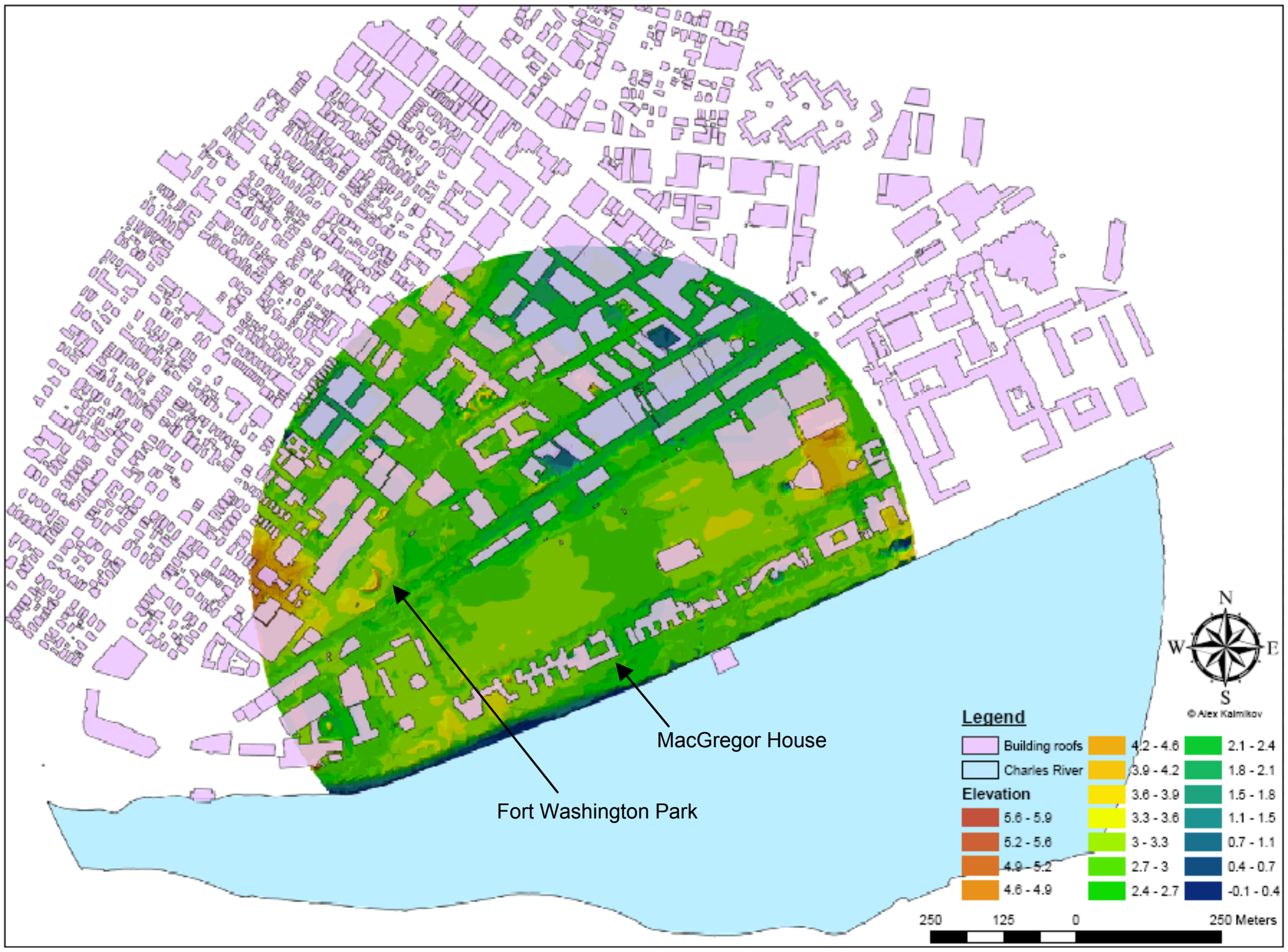

**Figure 2.1.** GIS map of MIT campus and the surroundings. Building roofs, river outlines and topography shown.

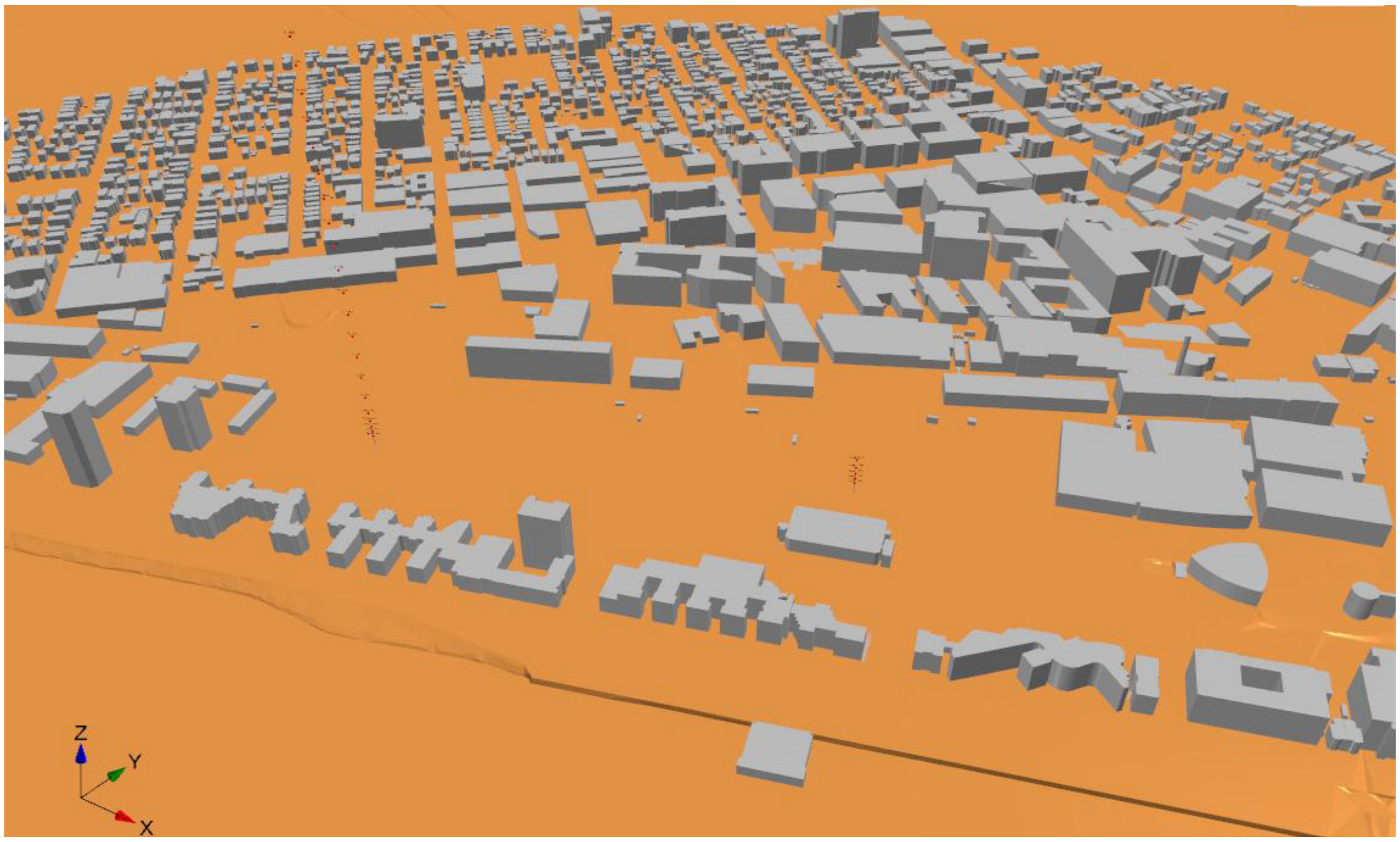

**Figure 2.2.** Three dimensional model of MIT campus geometry.

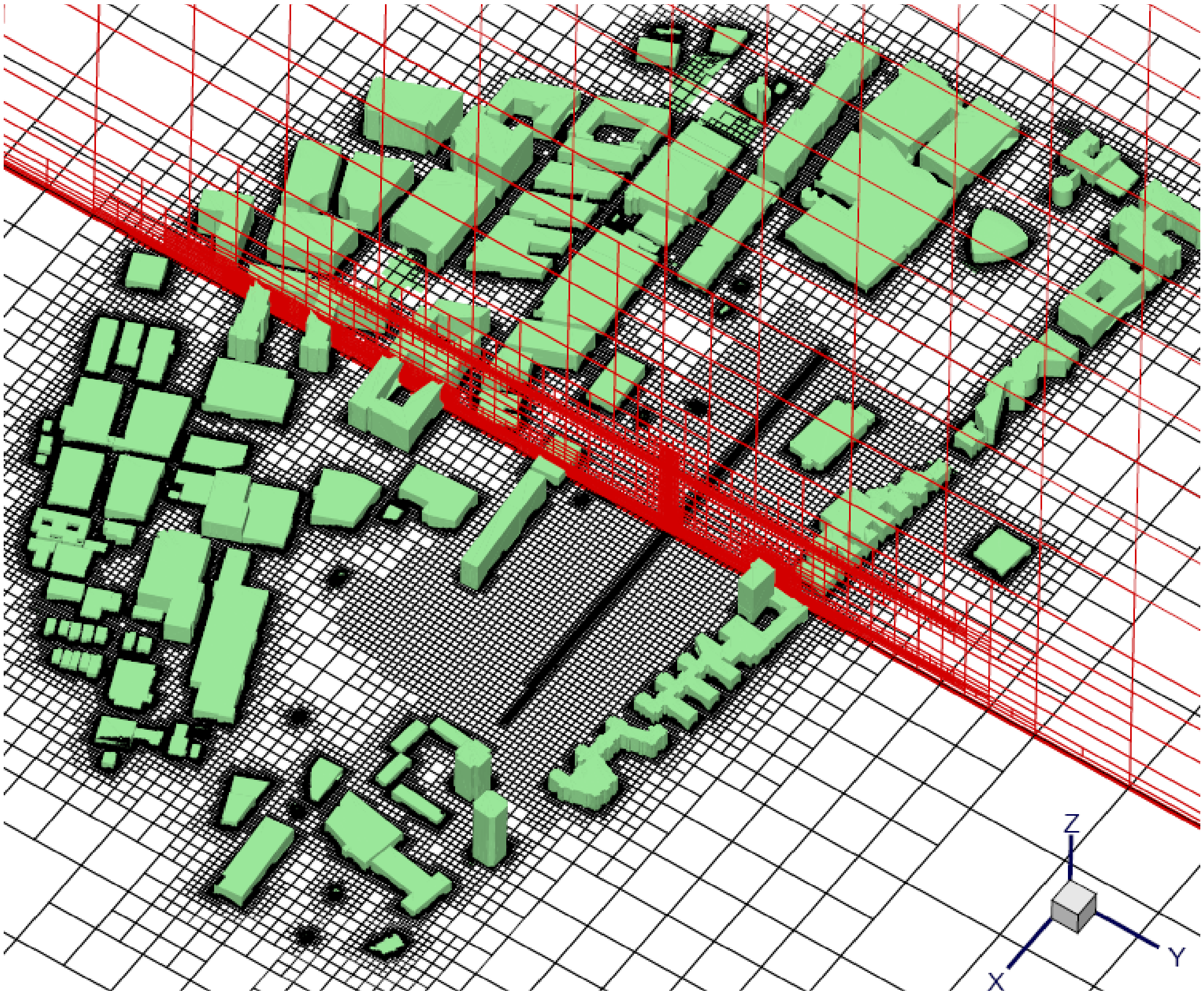

**Figure 2.3.** Adaptive meshing grid with maximal resolution 1 m. Horizontal and vertical cross sections shown.

# 3 MacGregor Wind Tunnel

## 3.1 The myth

The most common effect experienced by a pedestrian walking along Amherst Alley is the strong wind blowing along the street from the west direction (see also e.g. Wannaphahoon 2011). Near the corner of MacGregor House (Amherst and Fowler St. intersection) the wind can be so strong as to break umbrellas (H. Rebecca 2010) and flip over an incautious bicyclist (personal observation). This wind pattern appears to prevail under different weather conditions and is most strongly felt in the winter months (see also e.g. Balsley 1994). The common myth assumes that it is the unique architectural design of MacGregor House that causes this persistent wind acceleration, and although the exact mechanism is not known it is colloquially associated with a wind tunnel.

## 3.2 Wind micro-climatology analysis

To explain this phenomenon we first analyze the large-scale background wind conditions over the MIT campus. We derive wind statistics from the observations of the automated weather station on a mast on top of Green Building at the highest location on the MIT campus elevated about 99 m above the ground (Bicknell 1965). The directional statistics are shown in Figure 3.1 for 3 winter months (2010) and compared to the annual statistics over a 2 year period (2009-2010). It can be seen that the west-north-westerly direction prevails throughout the year but is particularly common in the winter. This prevailing wind direction corresponds to inflow from the open area to the north-west of the Brigg's Sports Field – around the Fort Washington Park (see Figures 2.1 and 2.2). As this flow approaches the MacGregor House it has a long fetch to accelerate near the ground due to the minimal roughness of this flat unobstructed terrain. Moreover, the strong wind current over the sports field has limited outlet from the basin generated by the buildings surrounding the field. The opening in the southern row of the buildings is located at Fowler St. just around the corner from MacGregor House. Thus, the geometric effect of the campus terrain is: first to accelerate the wind over the open areas and then to release the flow through the narrow opening just past the MacGregor House corner.

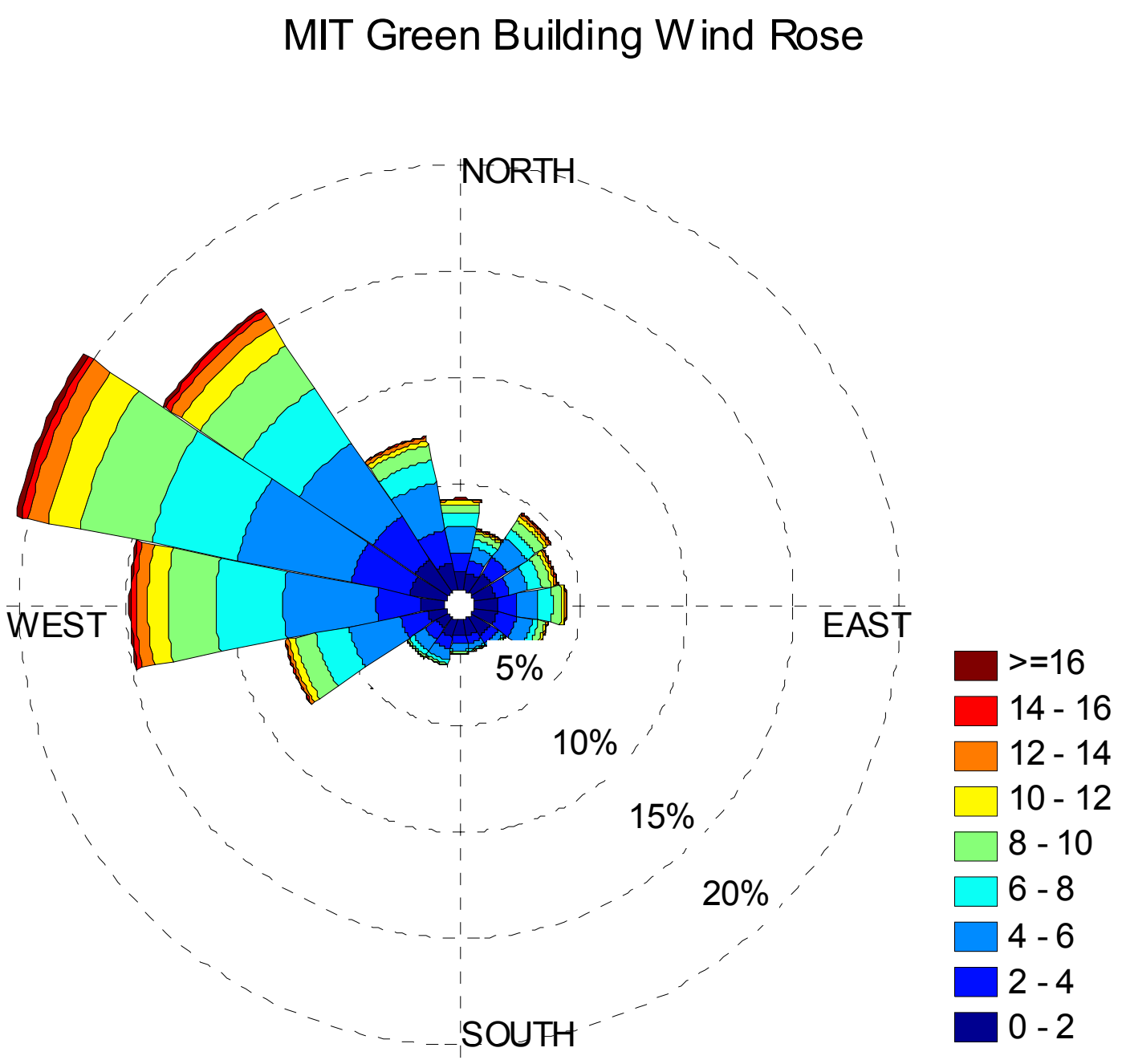

*(a.)*

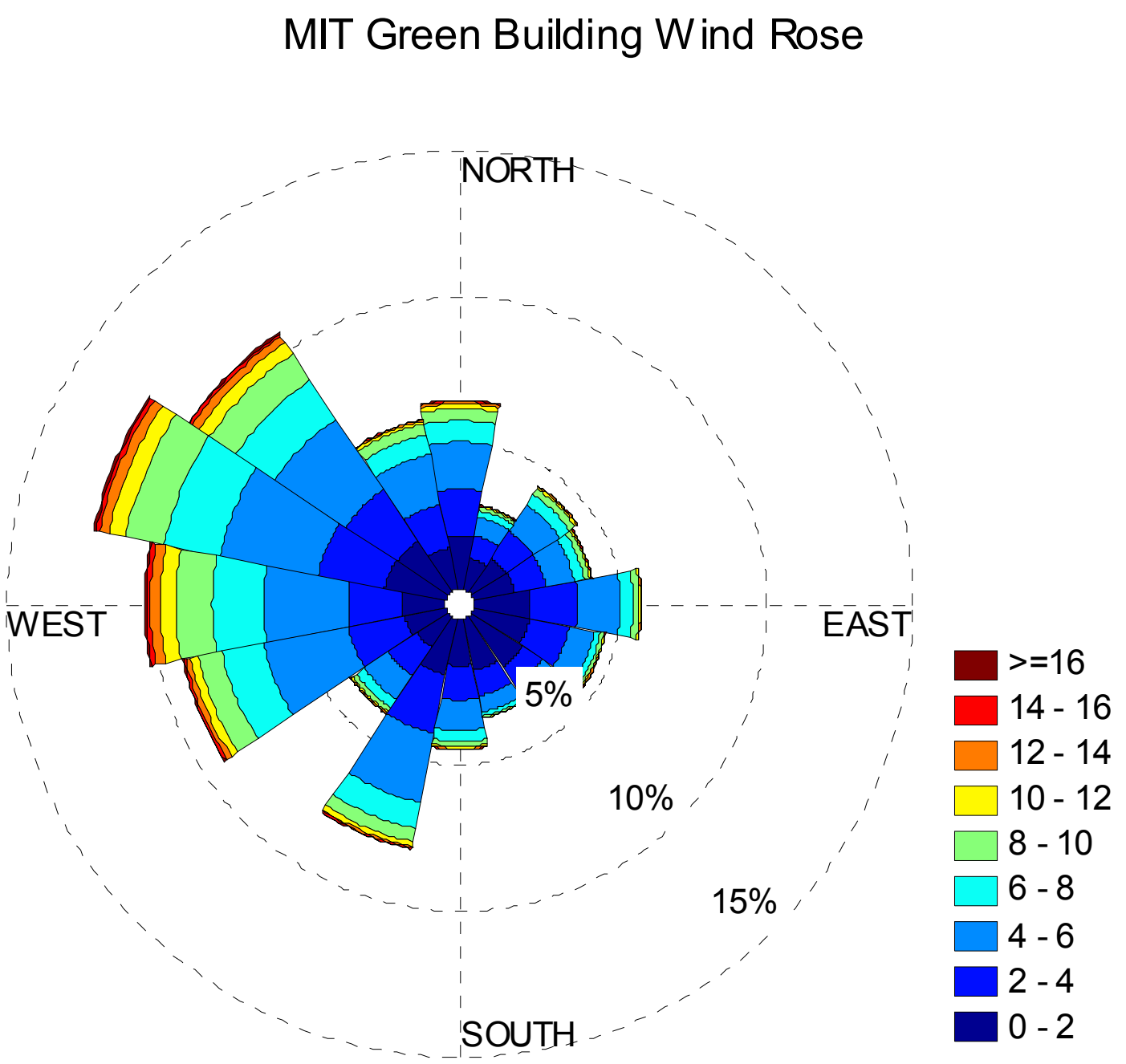

*(b.)*

**Figure 3.1.** Green Building wind roses – directional wind statistics, the radial dimension represents the frequency of wind occurrence in each of the directional sectors. Shown for: (a.) 3 winter months and (b.) for 2 years of data.

In the narrow passage between the buildings along Fowler St. the "Venturi effect" mechanism can be expected to further accelerate the wind. Although the original meaning of this mechanism[1] refers to a steady state acceleration of a confined flow forced incompressibly through a narrow cross-section, the Venturi effect is commonly used in wind engineering and urban aerodynamics to describe speed increase due to flow constriction in nonconfined flows (Blocken et al. 2008). Moreover, since the Venturi effect is used in design of actual wind tunnels (e.g. MIT's Wright Brothers Wind Tunnel) to accelerate the flow in the testing section, this association explains the naming of the MacGregor strong wind phenomenon - "wind tunnel."

## *3.3 CFD results*

We confirm the micro-climatological wind analysis with the CFD simulations as shown in the figures below. Figure 3.2 demonstrates the local acceleration effect for the prevailing large-scale flow direction – wind blowing from 280 degrees azimuth. Wind speed is shown in horizontal cross section 10 m above the ground, normalized by the reference speed at 100 m height. The highest wind speeds are concentrated around the north-eastern corner of MacGregor House, as well as behind the corner of Simmons Hall building and (to a lesser degree) between Burton-Conner House and the Tennis Bubble. A wide stream of high wind is seen accelerating diagonally across the sports field and impinging on the northern side of the Dorm Row buildings walls. The boundary layer on the upwind side of these buildings is thinner than downwind, the strong open field winds are closer to the dorms' walls on their northern side. Therefore, a pedestrian walking along Amherst Alley will be exposed to strong winds and would need to cross to the Memorial Dr. side of Dorm Row for shelter. Sensitivity to the background flow direction was qualitatively analyzed by comparing wind speed-up patterns for other prevailing directions: 270, 290, 300 (not shown). All these cases exhibit similar acceleration over the open area of the sports field extending to or reaching local maximum near MacGregor House.

[1] A closely related mechanism, frequently associated with Venturi effect, is the drop in pressure accompanying flow velocity increase in inviscid fluids, known as Bernoulli's principle.

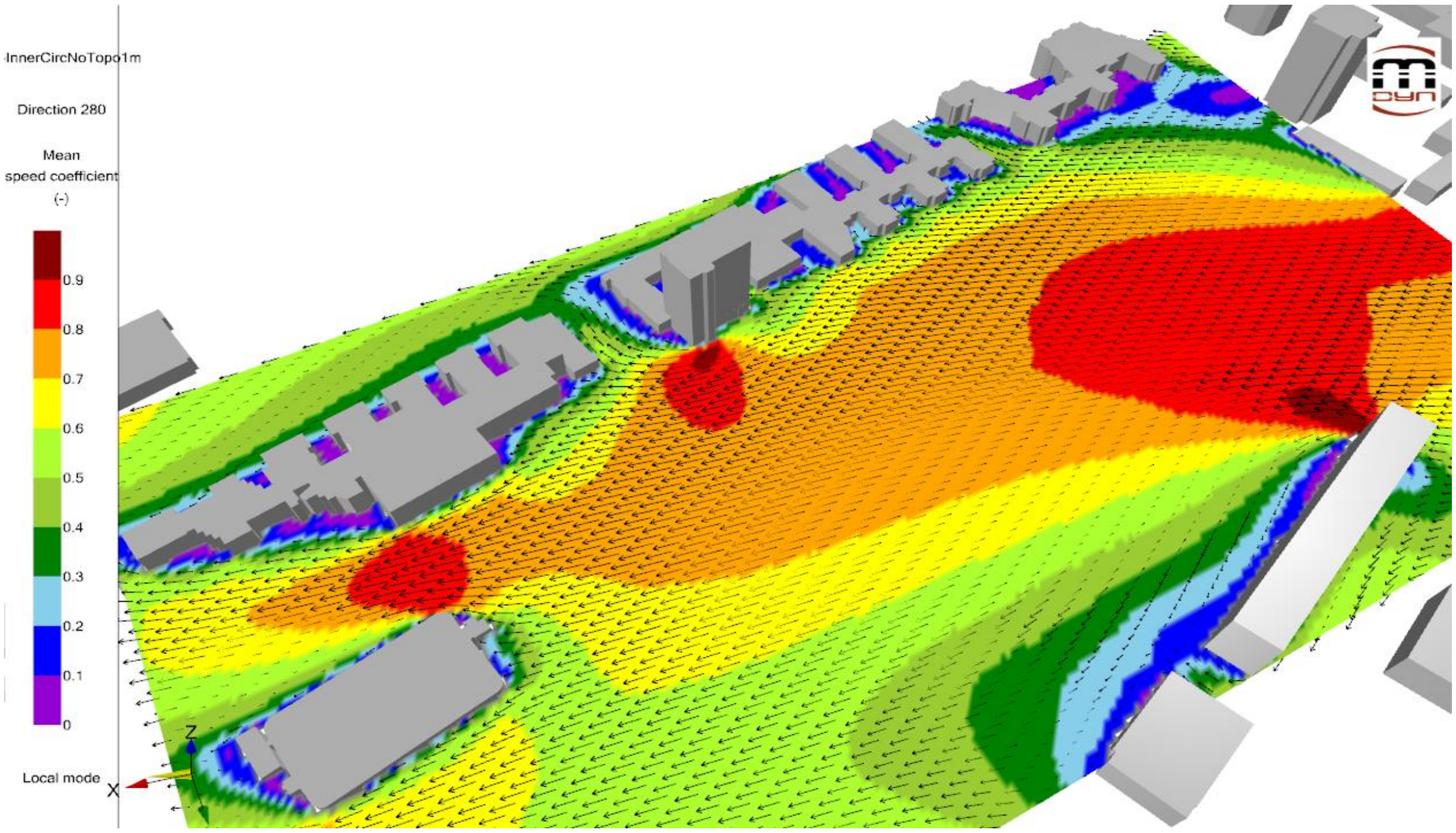

**Figure 3.2.** Wind speed map at 10 m at the prevailing climatological direction 280 degrees. Colors show the ratio of time average speeds to the reference wind speed averaged at 100 m height. Vectors show the magnitude and direction of the flow.

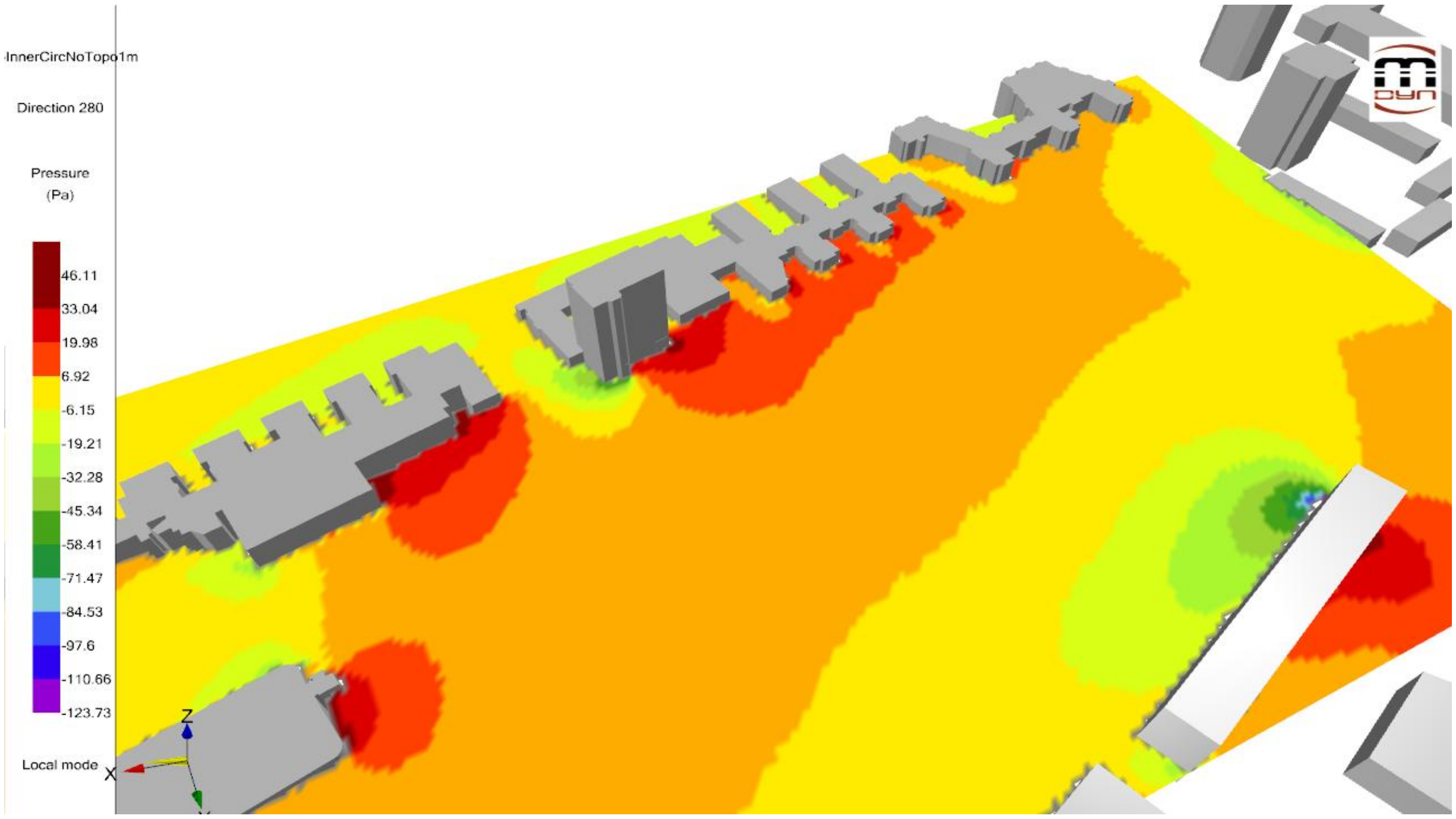

**Figure 3.3.** Pressure map at 10 m at the prevailing climatological direction 280 degrees.

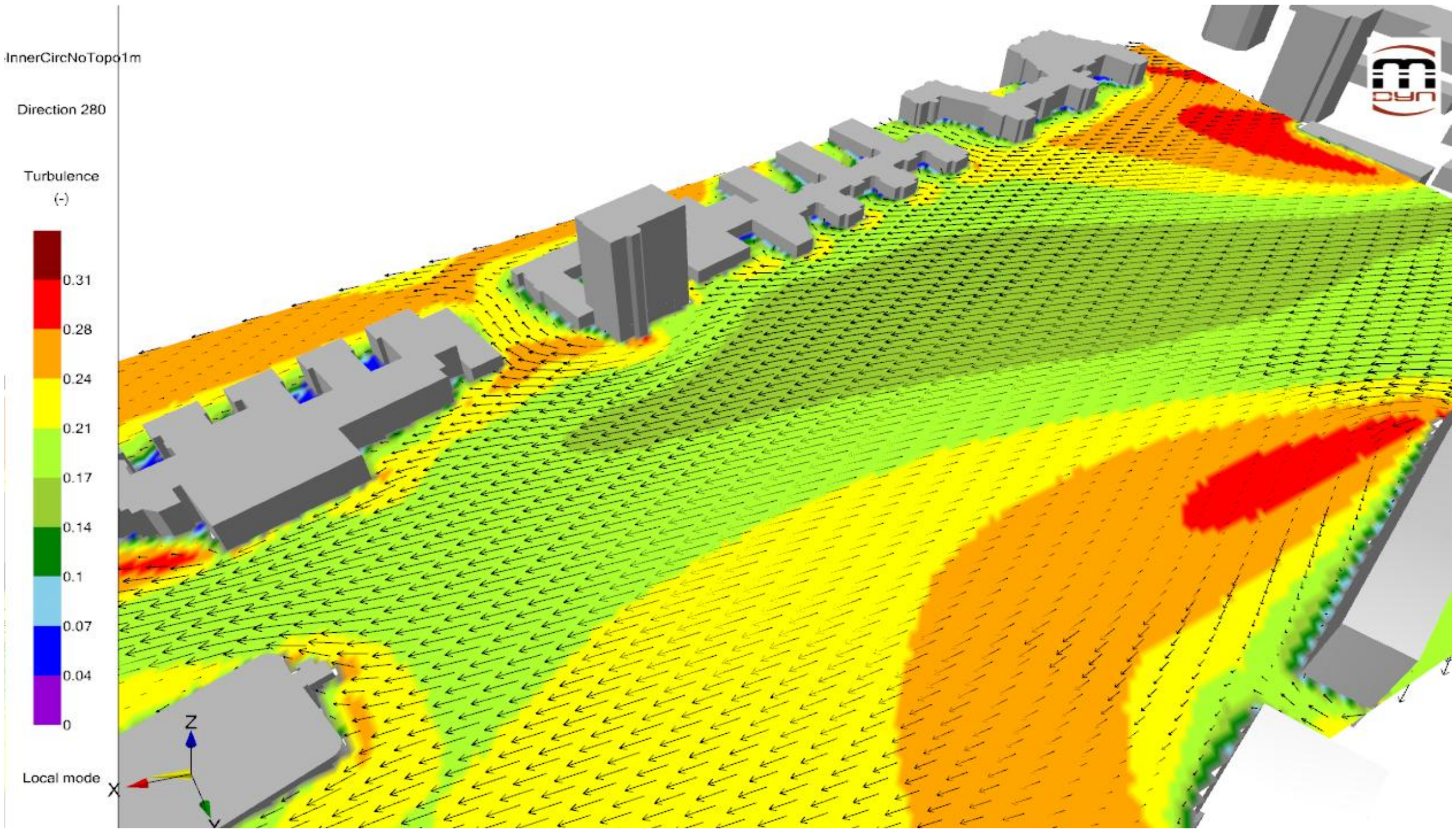

**Figure 3.4.** Turbulence map at 10 m at the prevailing climatological direction 280 degrees. Shown as square root of turbulent kinetic energy normalized by the average wind speed at 100 m height.

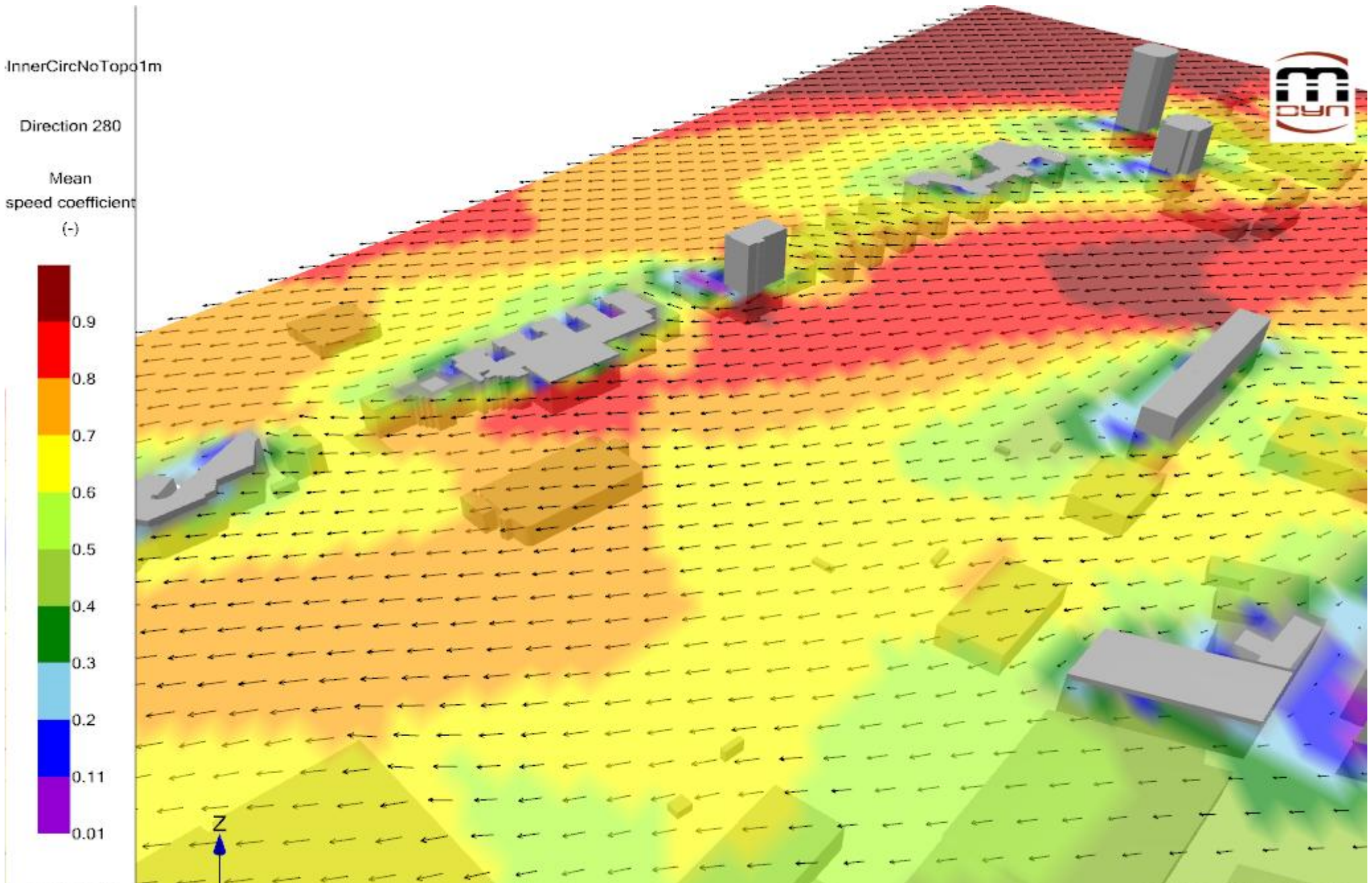

**Figure 3.5.** Wind speed map at 20 m at the prevailing climatological direction 280 degrees. Shown normalized by the reference speed value.

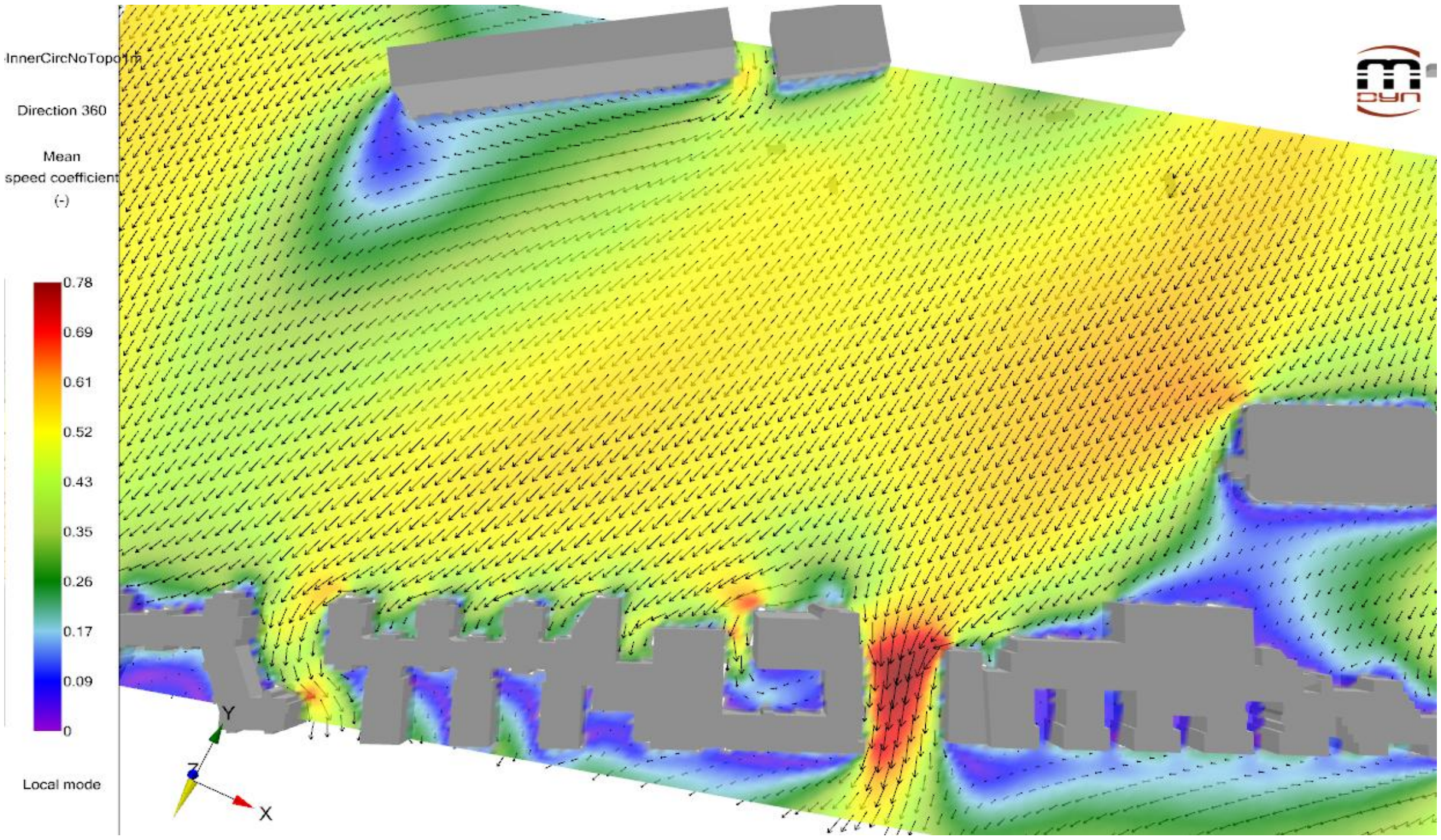

**Figure 3.6.** Wind speed map at 10 m at climatological direction 360 degrees. Shown normalized by the reference speed value.

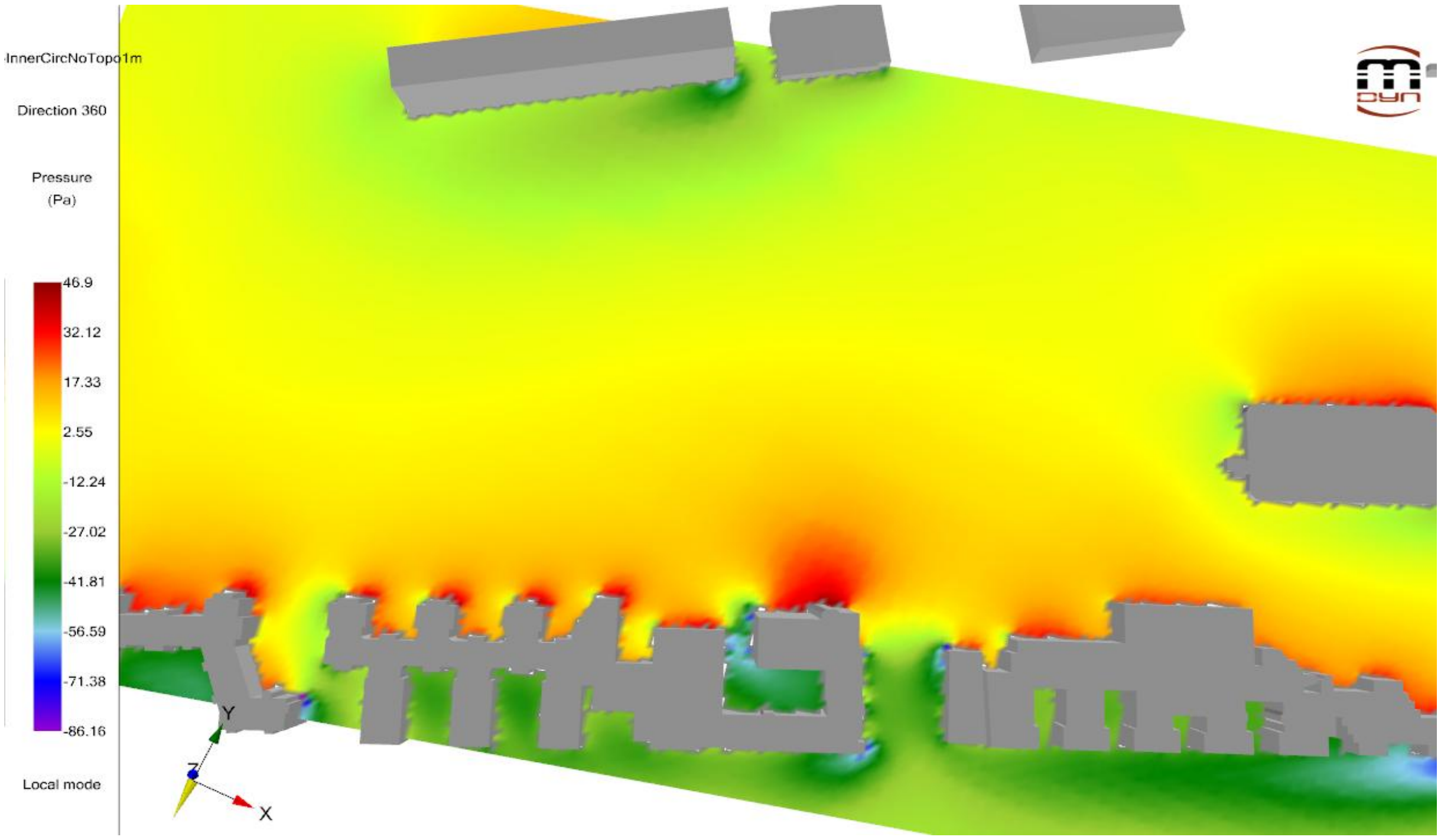

**Figure 3.7.** Pressure map at 10 m at climatological direction 360 degrees.

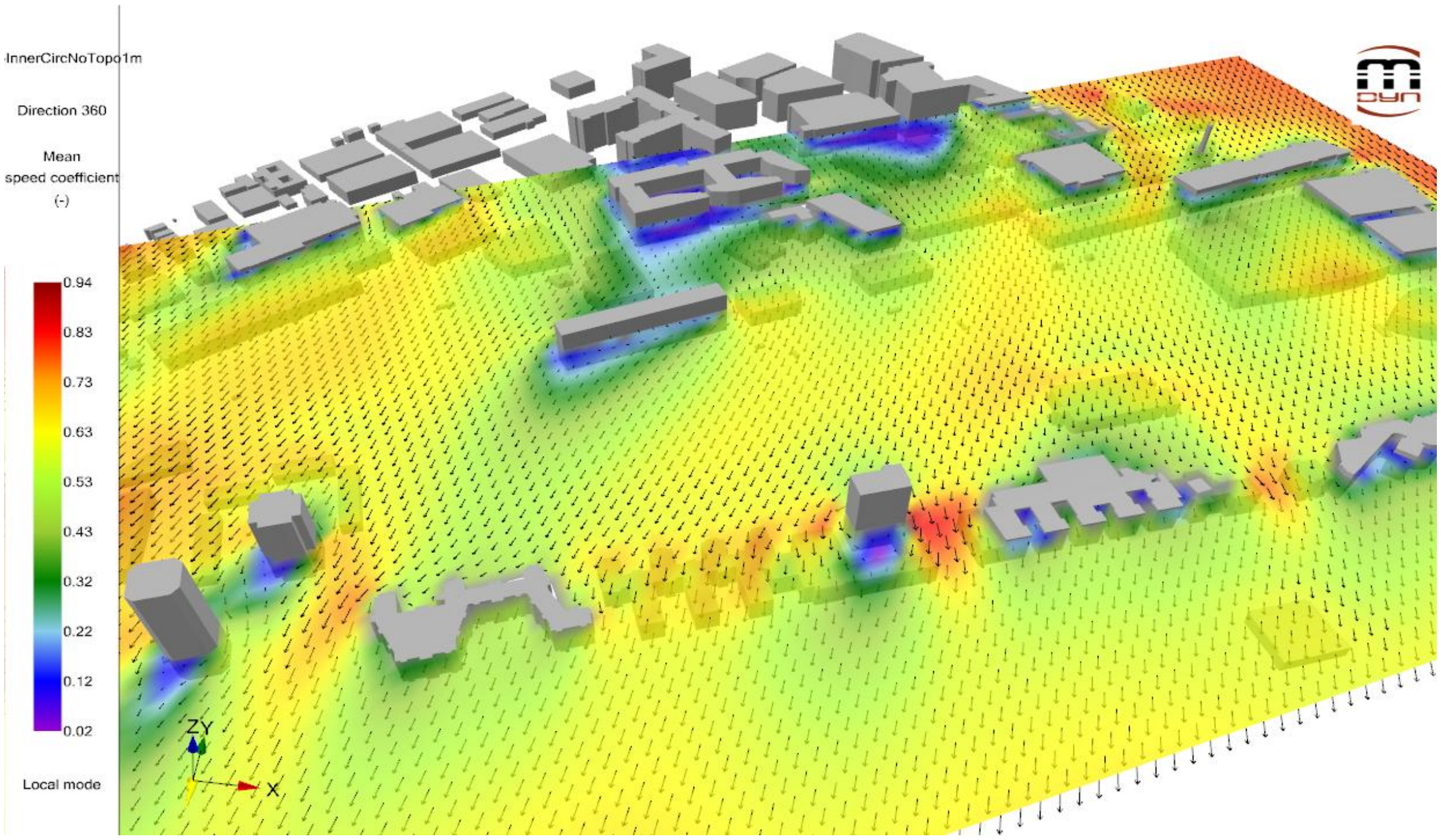

**Figure 3.8.** Wind speed map at 20 m at climatological direction 360 degrees. Shown normalized by the reference speed value.

The pressure map is shown in Figure 3.3 in terms of an average perturbation to the mean ambient pressure value. We note the strong pressure gradient around the north-eastern corner of MacGregor House – a drop of 100 Pa (1 mbar). It is responsible for the peak acceleration of the flow. The mechanism of this pressure gradient can be explained by combination of the large stagnation pressure associated with the deceleration of the impinging flow on the northern side of the Dorm Row buildings (Bernoulli's principle), with pressure deficit behind the corner of MacGregor House in the urban canyon of Fowler St. Large pressure drop is located also behind Simmons Hall and a moderate drop between the Burton-Conner House and the Tennis Bubble. In all these cases the pressure gradient force balances the turbulent friction, explaining the localized high winds in the steady state.

The turbulence intensity, defined here as square root of turbulent kinetic energy normalized by the reference wind speed, is shown in Figure 3.4. The low values over the sports field highlight the relatively weak turbulent friction as wind accelerates in the open area from Fort Washington Park to MacGregor House of the sports field. The corner of the MacGregor House experiences higher turbulence levels as well as other regions of high pressure gradients (Figure 3.3).

For a comparison we display the wind speeds at 20 m over the ground (Figure 3.5). The elevated wind jet developed over the sports field appears as a wide continuous wind current where wind speed is twice as large as in the surrounding areas. This jet blows directly at the north-eastern corner of MacGregor House, demonstrating that high winds persist also on higher floor levels. We can also note the homogeneous high winds at the very top of the figure developing over the open area of Charles river, where absence of urban obstacles allows wind to accelerate to maximum speed.

The analysis presented above provides a different explanation of the high winds at the base of MacGregor House than what is implied by the name "MacGregor Wind Tunnel." These are the prevailing wind conditions in the winter and also most frequent annually. Nonetheless, occasionally the background larger-scale wind blows from north or south and a different wind flow pattern is established around MacGregor House, which much closer matches the intuitive expectation of "MacGregor Wind Tunnel" flow. Under the conditions of northern background winds (Figure 3.6) the flow has shorter acceleration fetch over the open area of the sports field and is channeled through the gap at Fowler St. without changing the flow direction. A clear "wind tunnel" (Venturi) pattern is generated in the channel with the speeds at the jet about 50% higher than upwind before the channel. The corresponding pressure map is shown in Figure 3.7, highlighting a strong gradient of pressure across the wind tunnel entry. The location of the maximum wind speed correlates well with the maximum pressure gradient, explaining the mechanism of the wind tunnel flow. Figure 3.8 adds a higher level view of the channel jet 20 m above the ground. The coherent three-dimensional structure of Venturi flow is confirmed up to height where the gap between the buildings ends. This CFD simulation of local wind acceleration in the passage between parallel buildings supports the public perception of urban wind tunnel flow. This result may be compared to recent findings with a modified CFD model that Venturi effect between parallel buildings may be rather weak, with flow intensification limited to the pedestrian levels only and no coherent thee-dimensional jet (Blocken et al. 2007).

### 3.4 CFD micro-climatology

To complete the analysis we calculate the climatological weighted average of wind speeds to produce mean seasonal and annual wind maps. Normalized directional wind speed maps, such as in Figure 3.2, were calculated with the CFD model for each of the directions in wind rose statistics (Figure 3.1). Then, a directional weighted average of the wind maps was calculated based on the background wind climatology from the Green Building observations. The resulting seasonal winter wind map is shown in Figure 3.9. The open area of the sports field is clearly dominated by high winds. Interestingly, these high winds extend to the north-eastern corner of MacGregor House. This is a unique feature among the Amherst Alley dorms, which are exposed to weaker winds close to the buildings. A similar area of average high winds very close to the building is estimated also near the south-western Simmons Hall corner. Strong winds are also expected between Burton-Conner House and the Tennis Bubble. In contrast to these high wind zones, the urban channel along Fowler St. experiences on average weaker wind speeds. This does not contradict the expectation of extreme wind conditions due to the Venturi effect in the channel (Figure 3.6) on atypical winter days. Moreover, the understanding of directional flow patterns for different background flow conditions enables micro-meteorological urban flow prediction based on larger scale models or observations.

A similar micro-climatological average wind map for the 2 year statistics is shown in Figure 3.10. It is seen that the acceleration effect at the corner of MacGregor House is milder, understandably due to the more frequent occurrence of background winds from other directions (Figure 3.1). Moreover, the average wind speeds are weaker all over the domain, as expected due to the slower background flow in the other seasons. As both winter and annual wind maps show - the coherent wind tunnel flow structure (Figure 3.6) is not a common flow feature on average. It is the unique alignment of the prevailing wind direction with the long unobstructed acceleration path which explains better the high winds experienced at the base of MacGregor House. One may then say that this is the true "MacGregor wind tunnel." In any case, based on the presented analysis these high winds do not appear to be dependent on the architecture of MacGregor House itself.

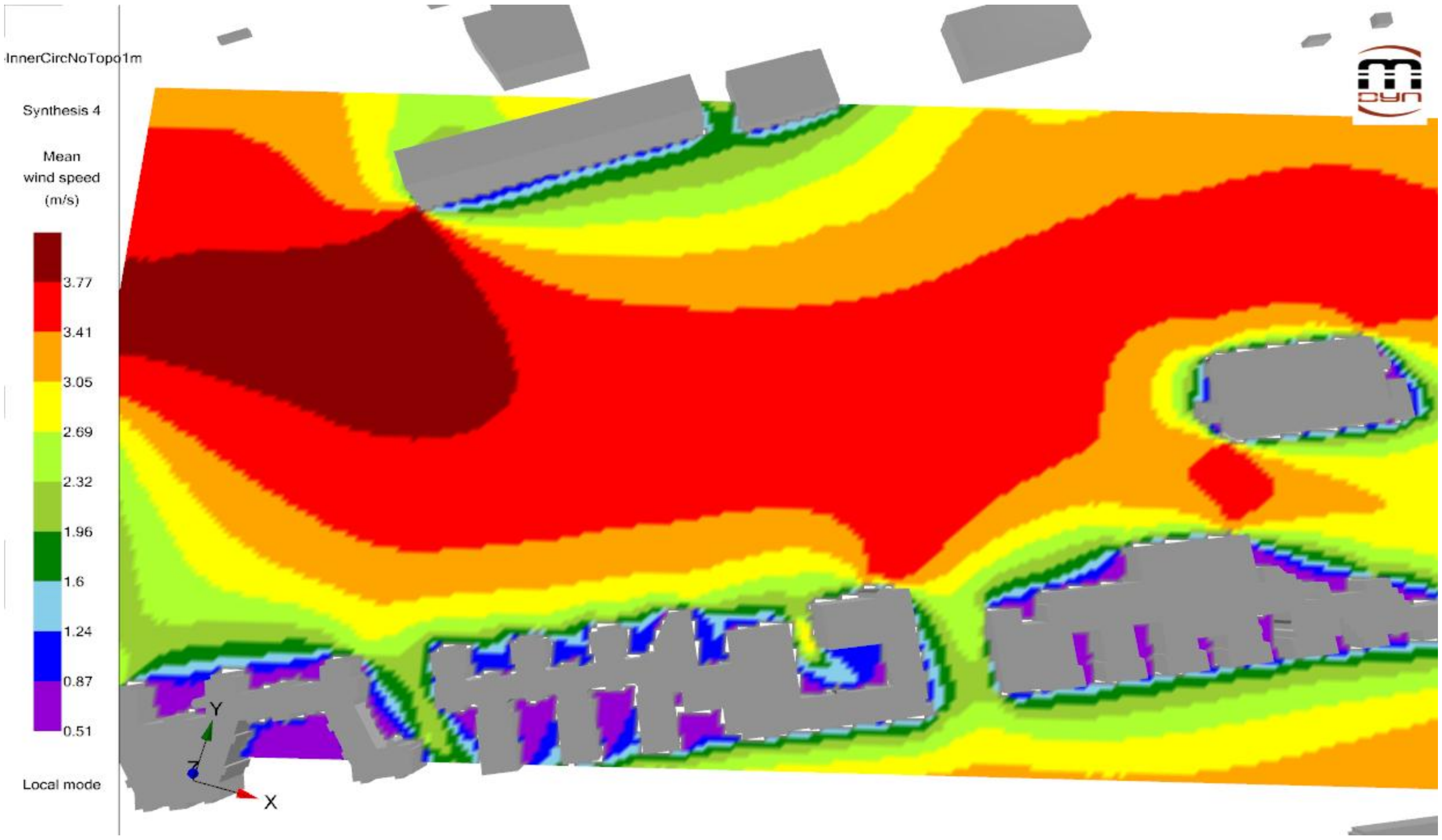

**Figure 3.9.** Composite climatological wind speed average for 3 winter months.

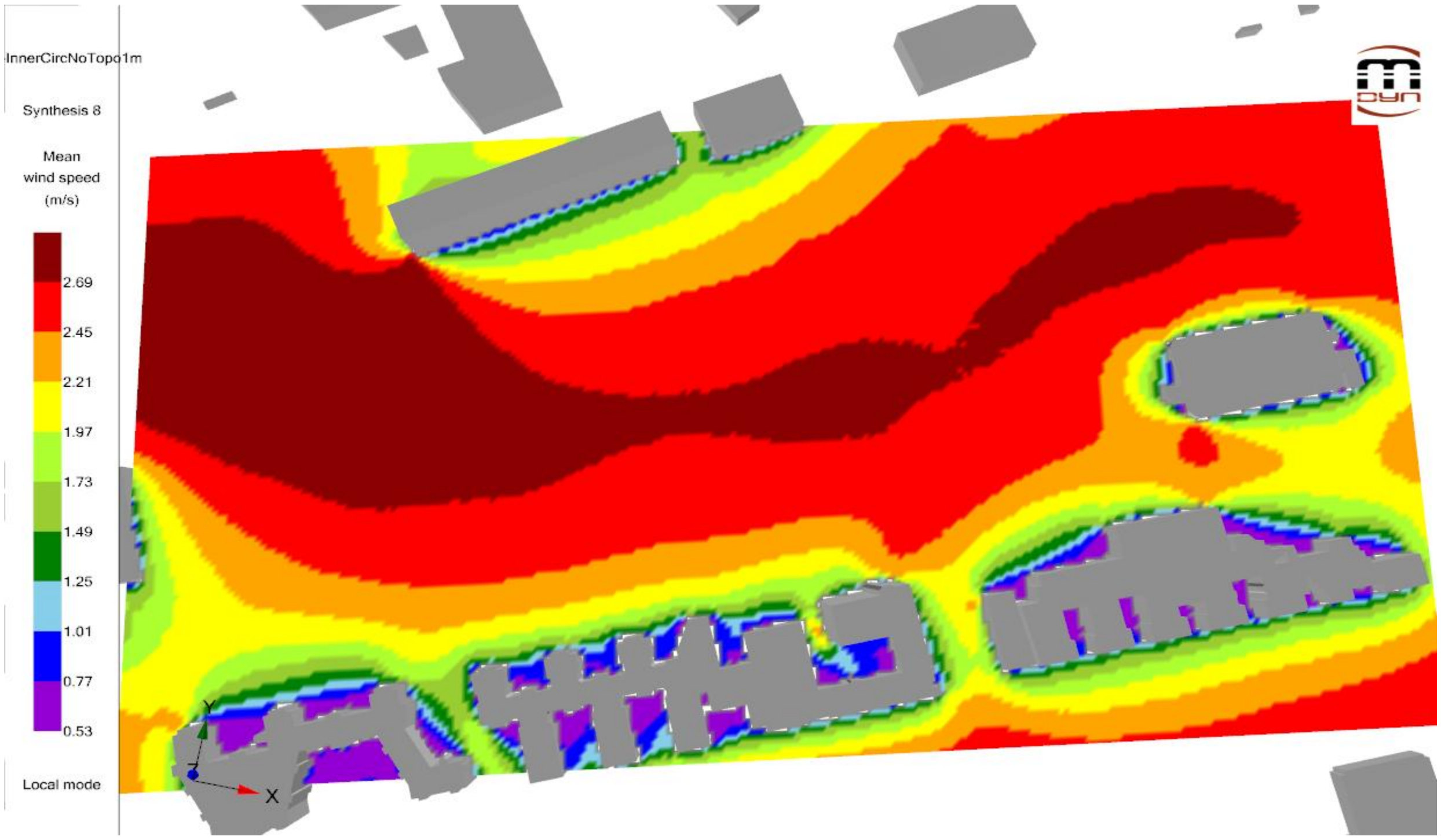

**Figure 3.10.** Composite climatological wind speed average for 2 years

# 4 Green Building and The Big Sail

## 4.1 The myth

Strong winds around Green Building attracted public attention soon after its construction. Doors in the ground floor lobby were difficult to open on windy days and just two months after the building was dedicated the door on the northern side was blown in by 35 mph December winds (Burke and McCaffrey 1965). It was then boarded up, while detailed wind studies were made to design a solution to building entry. Soon after, two windows, one on the 15th floor and the other higher up, blew out and fell to the ground. A comprehensive experimental study was carried out in Wright Brothers Wind Tunnel at MIT, which confirmed an acceleration effect on the wind at the base of the building (Bicknell 1965, 1966). Different solutions were tested, such as a complete enclosure of the ground floor and different revolving doors designs, including the one that the building has today. The Big Sail stabile was installed in the McDermott Court in front of Green Building in 1966, the same year when the wind tunnel reports were completed. In fact, the stabile was also tested in the wind tunnel, but for its own stability only (Gleitzman 2006, Bourzac 2006). But the coincidence of both the series of wind tunnel studies and the solution to Green Building's windy doors problem may have lead to the birth of the myth.

## 4.2 Wind tunnel experiments

Aerodynamic model tests of wind flow around Green Building were carried out in the Wright Brothers Wind Tunnel at MIT (Bicknell 1965, 1966). A 1/96 scaled model of the 277 feet high tower and the surrounding buildings (Figure 4.1) was rotated at different angles to incoming flow in the tunnel. An artificial inflow shear layer was created with a non-uniform barrier of tapered pickets (Figure 4.1) to model the gradual increase of wind speeds above the ground (Bicknell 1965). Flow velocities were measured with hot-wire anemometers, pressure taps on the buildings and a pitot tube at the height corresponding to the real building's rooftop anemometer. Results of velocity surveys were compiled manually to produce wind maps normalized by the rooftop velocity (Figure 4.2). The tedious procedure, low resolution results and lack of flow direction information led to adoption of an alternative technique using deflection type indicators (Bicknell 1966) made of wrapped pieces of fine solder, called "men" (also "little men", Durgin 1992). Placed around the building to model 6 ft tall pedestrians, the indicators would bend under the force of wind and their photographs were used to record the flow patterns (Figures 4.3 and 4.4). The level of detail and quality of visualization of these classical techniques of wind engineering can be compared to the power of CFD modeling, as illustrated in the next section. Nonetheless, these experiments provided an invaluable insight into wind patterns around Green Building and enabled development of a solution to the problem of too strong winds, which allows the building to operate up to this day.

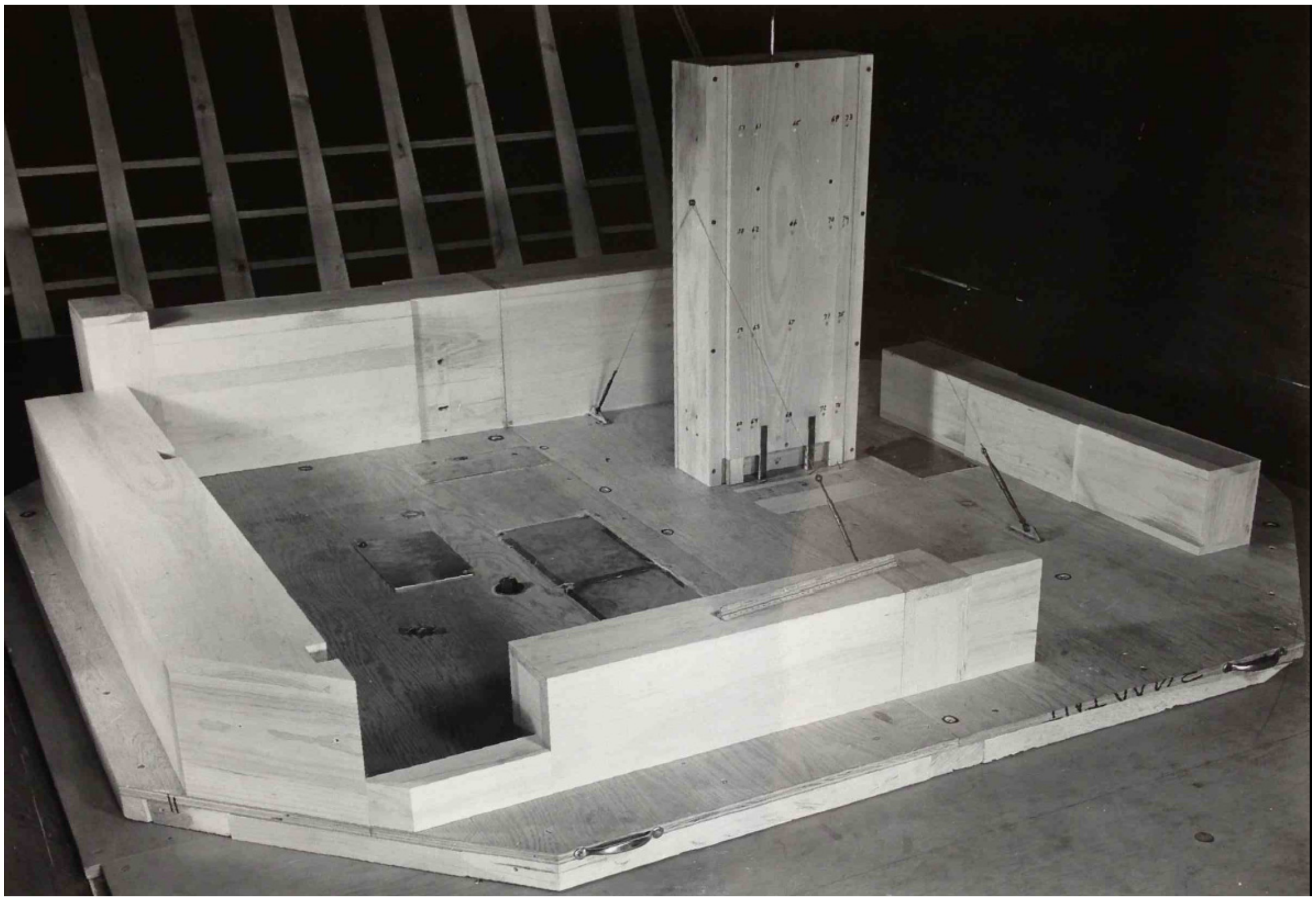

**Figure 4.1** Scaled model of Green Building and the surrounding building set up in wind tunnel. Tapered picket barrier, used to simulate the boundary layer shear, is seen in the background. Source: Wright Brothers Wind Tunnel Report 1027, Courtesy of MIT Libraries, Institute Archives & Special Collections, Cambridge, Massachusetts, Massachusetts Institute of Technology, Wright Brothers Wind Tunnel records (AC144).

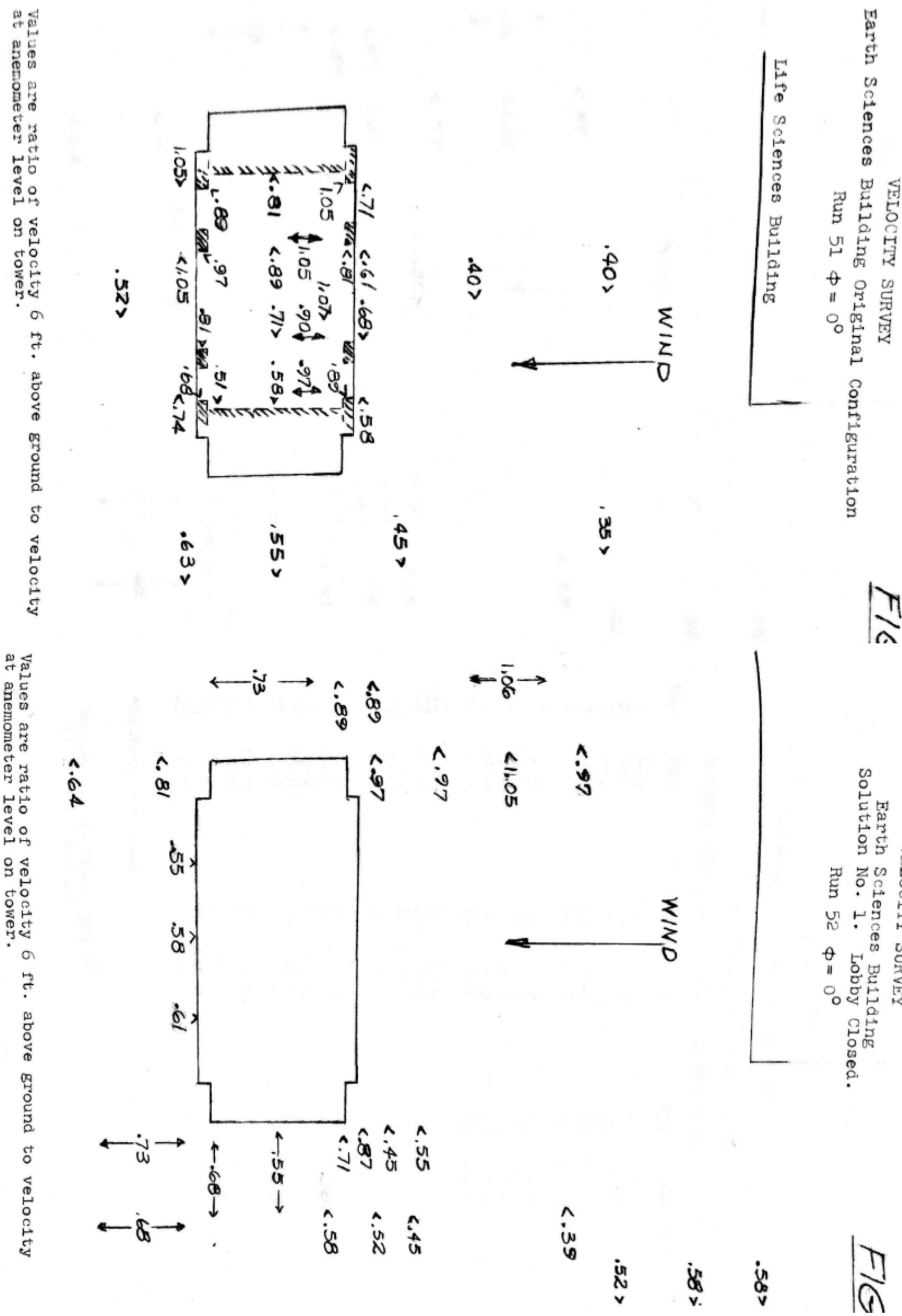

**Figure 4.2** Experimental wind maps compiled manually from individual flow sensors. Upper map shows internal flow acceleration in open lobby, lower map shows wind speeds around the building. Only flow velocity is shown, without directional information. Source: Wright Brothers Wind Tunnel Report 1027, Courtesy of MIT Libraries, Institute Archives & Special Collections, Cambridge, Massachusetts, Massachusetts Institute of Technology, Wright Brothers Wind Tunnel records (AC144).

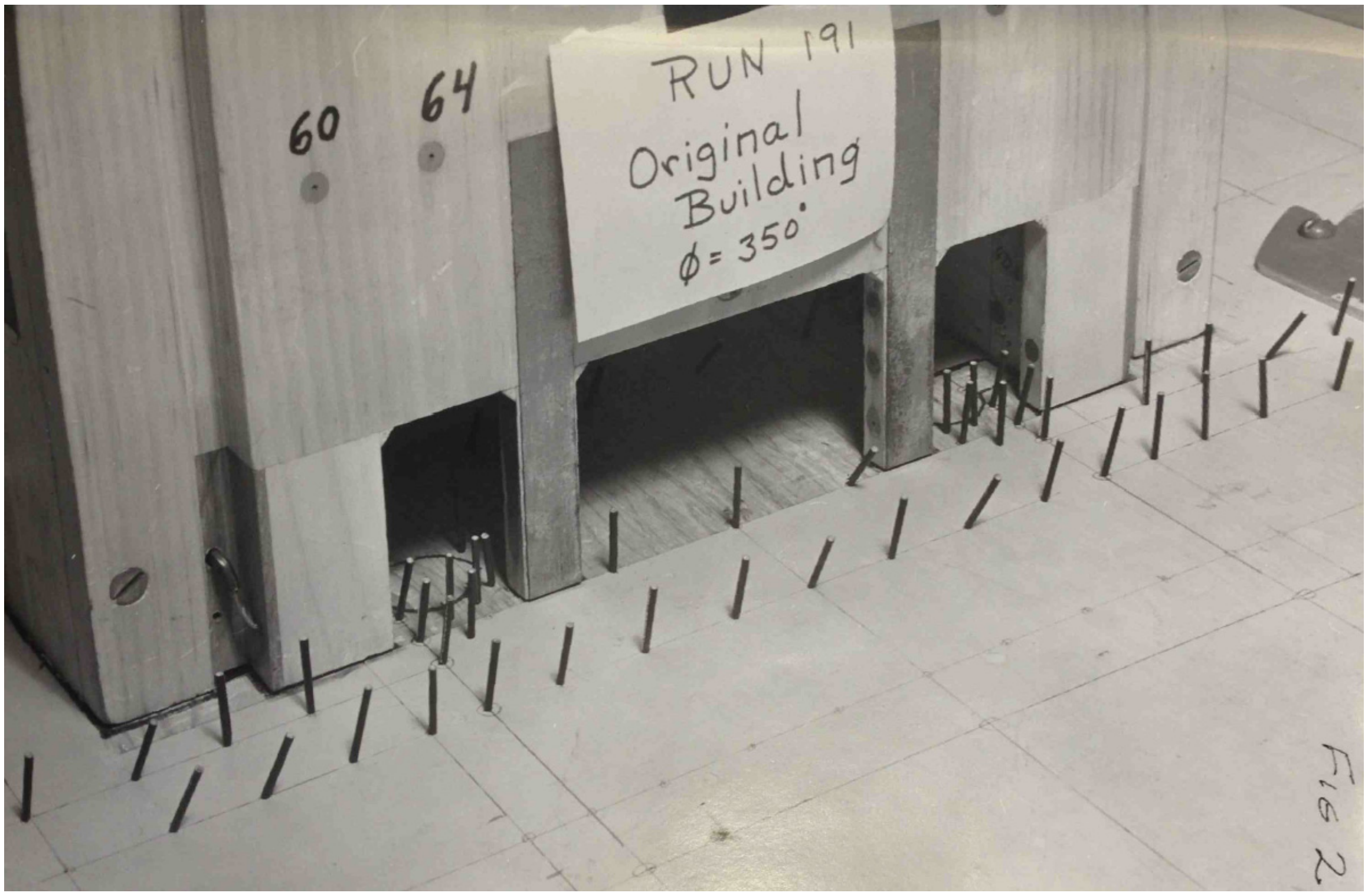

**Figure 4.3** Deflection type indicators, also know as "little men", modeling wind effect on 6 ft tall pedestrians. Source: Wright Brothers Wind Tunnel Report 1029, Courtesy of MIT Libraries, Institute Archives & Special Collections, Cambridge, Massachusetts, Massachusetts Institute of Technology, Wright Brothers Wind Tunnel records (AC144). 

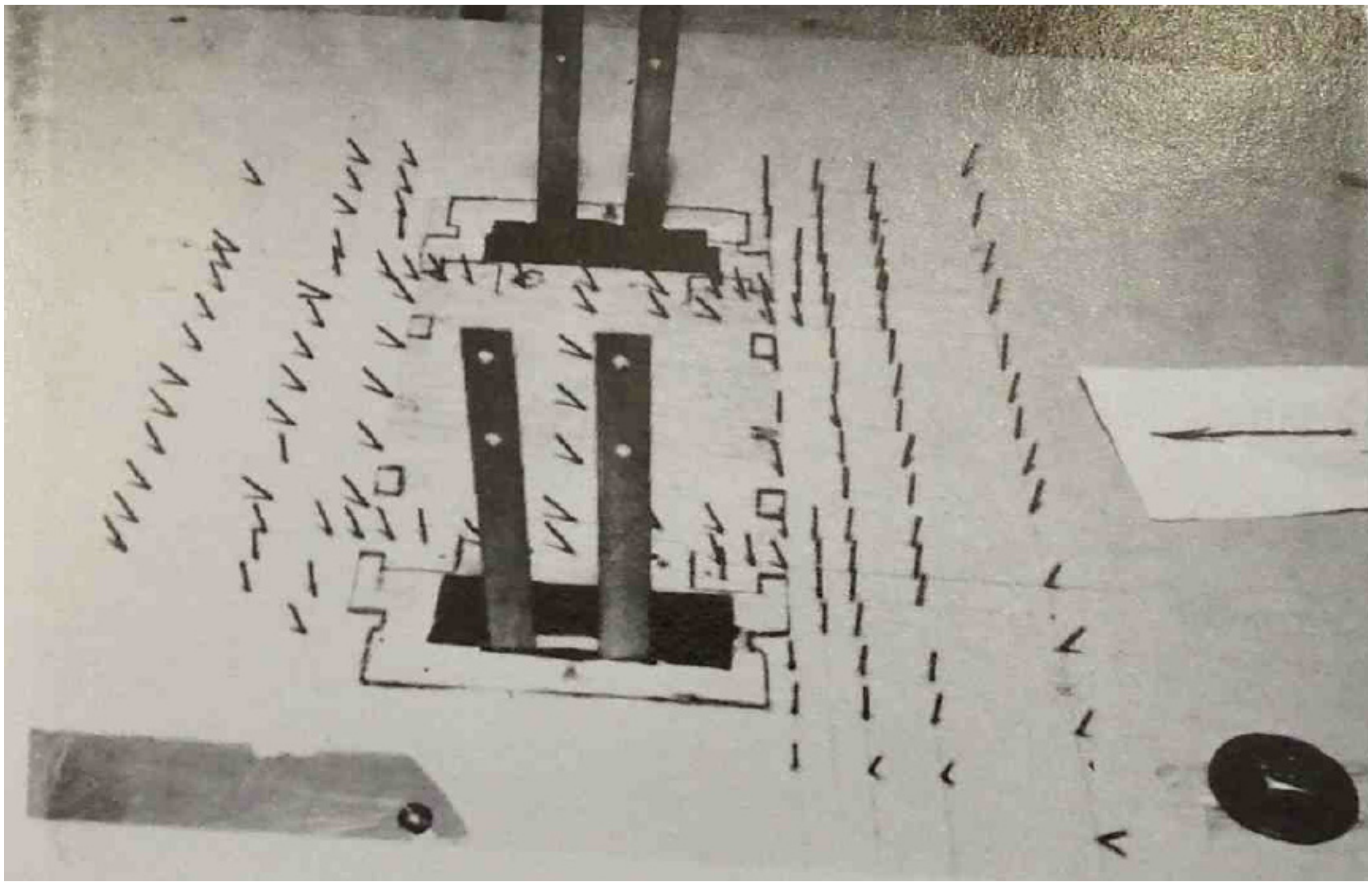

**Figure 4.4** Directional wind map, shown for background inflow direction marked by the arrow. The orientation of indicator's deflection shows the direction of local flow. The angle of the deflection shows the velocity.

Source: Wright Brothers Wind Tunnel Report 1028, Courtesy of MIT Libraries, Institute Archives & Special Collections, Cambridge, Massachusetts, Massachusetts Institute of Technology, Wright Brothers Wind Tunnel records (AC144). 

## 4.3 CFD results

To resolve the controversy surrounding the installation of "The Big Sail" stabile, its effects on the winds at the base of Green Building are resolved with the CFD method. Southern wind flow is simulated with the 40-foot tall stabile and without it (Figure 4.5). Four simulation configurations were examined: with the stabile modeled by a 40-foot tall cylinder with medium porosity (volumetric drag coefficient $C_D$=5), low porosity (volumetric drag coefficient $C_D$=9), as solid impermeable building, and without the stabile. The Green Building was modeled manually based on the GIS data and the floor plan from MIT Facilities. The external revolving door assemblies are not modeled to simulate the effects of the strong winds before their installation (Bicknell 1966).

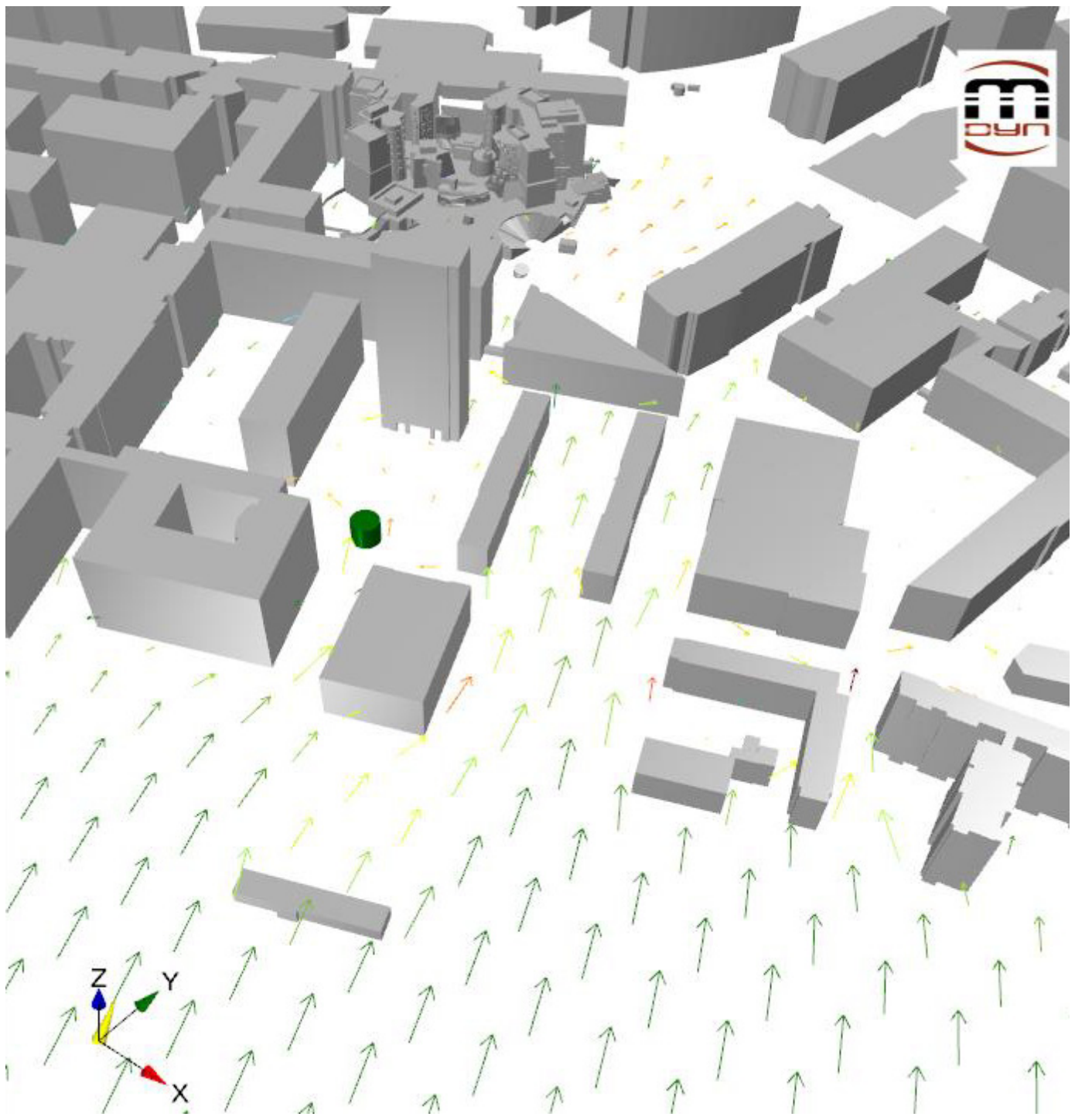

**Figure 4.5.** Upwind boundary condition - horisontaly uniform southern wind at azimuth 155 degrees. Shown at elevation 10 m. The big Sail stabile is shown with green cylinder.

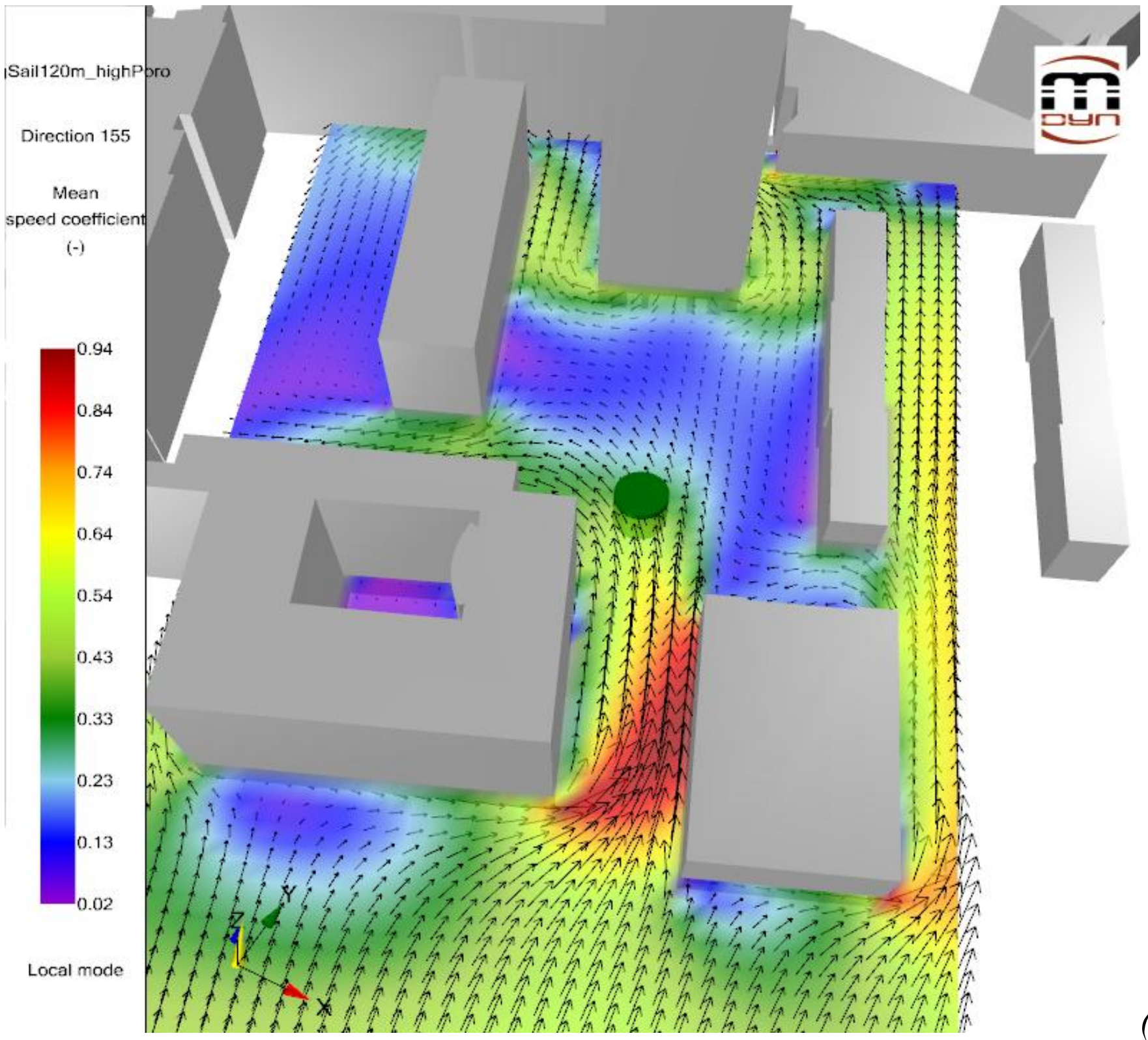

*(a)*.

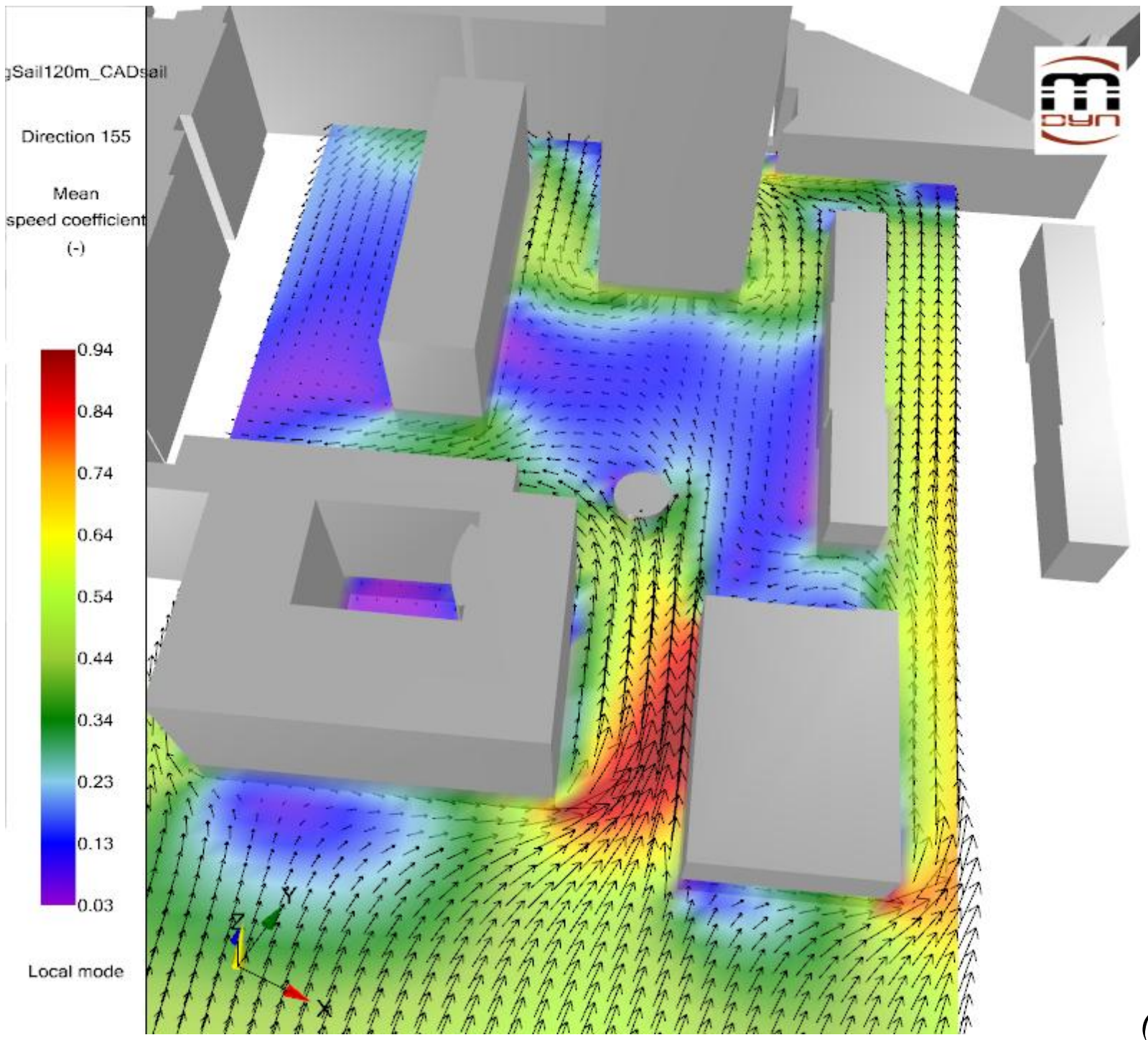

*(b)*.

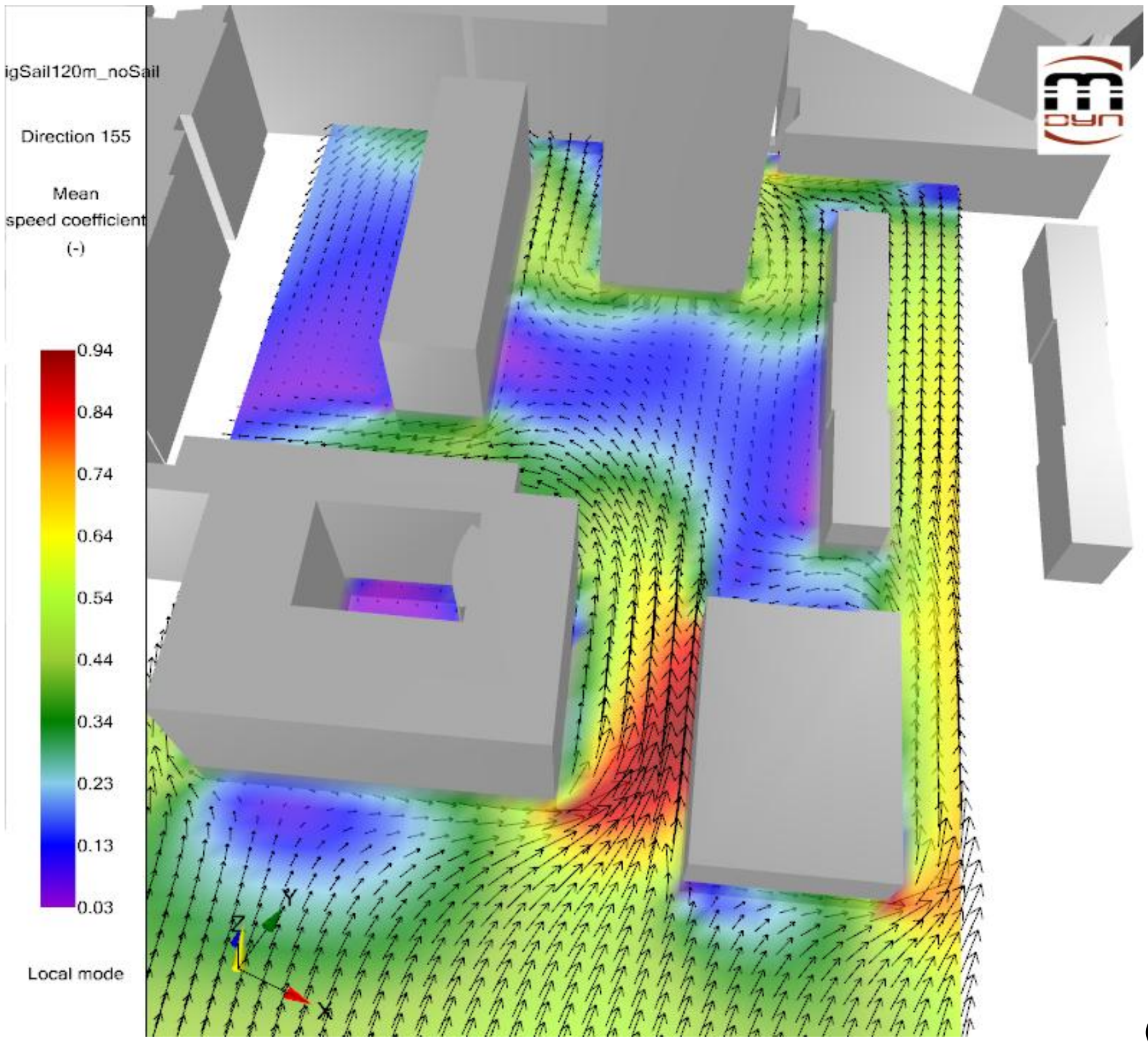

*(c)*.

**Figure 4.6.** Wind map at elevation 10 m, color shades visualize the wind speed. Cases shown: *(a)* - low porosity permeable stabile, *(b)* - impermeable solid stabile, *(c)* - no stabile.

The horizontal wind maps at elevation 10 m (Figure 4.6) show that the southern wind flow accelerates by a factor of 2 as it is channeled between Walker Memorial and Building 14. However, after the jet exits the tunnel, winds appear to slow down dramatically, up to a factor of 5 below the incoming wind. The flow is deflected to the left (westward, around Building 14) and the open area in front of the Green Building - the McDermott Court, experiences only very weak winds (blue shades in Figure 4.6). This is also true at elevation 2 m over the ground (see Figure 4.7). As it can be clearly seen in Figure 4.7 and Figure 4.8, the local flow around the stabile is quite different between the solid impermeable and porous cases. The flow is deflected around the solid cylinder but permeates through the porous cylinder while slowing down. Nevertheless, this difference is very localized and the large scale flow pattern is very similar among all 4 cases - low and medium porosity, solid and no stabile. (The medium porosity case is not shown in the figures.)

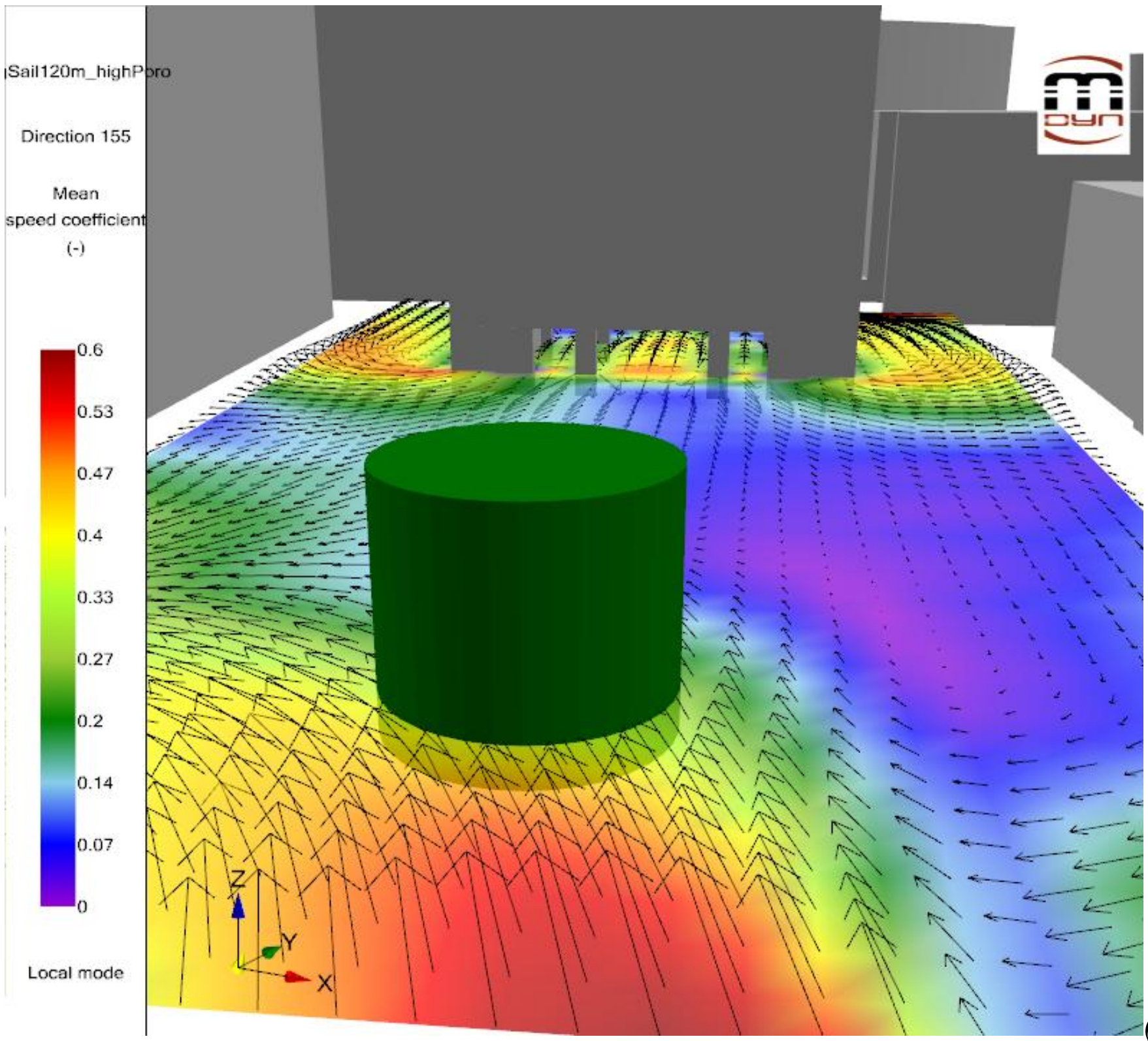
(a).

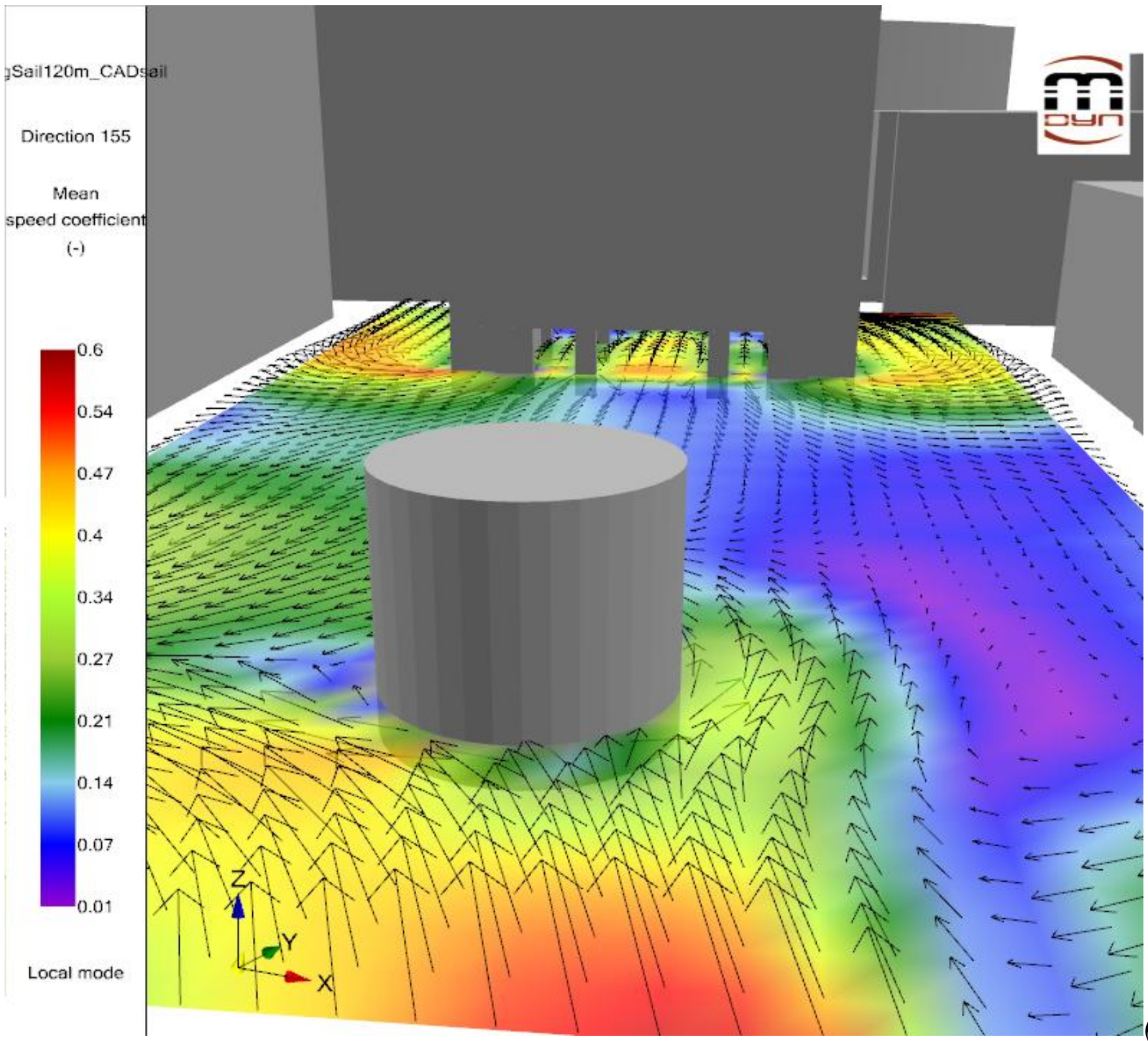
(b).

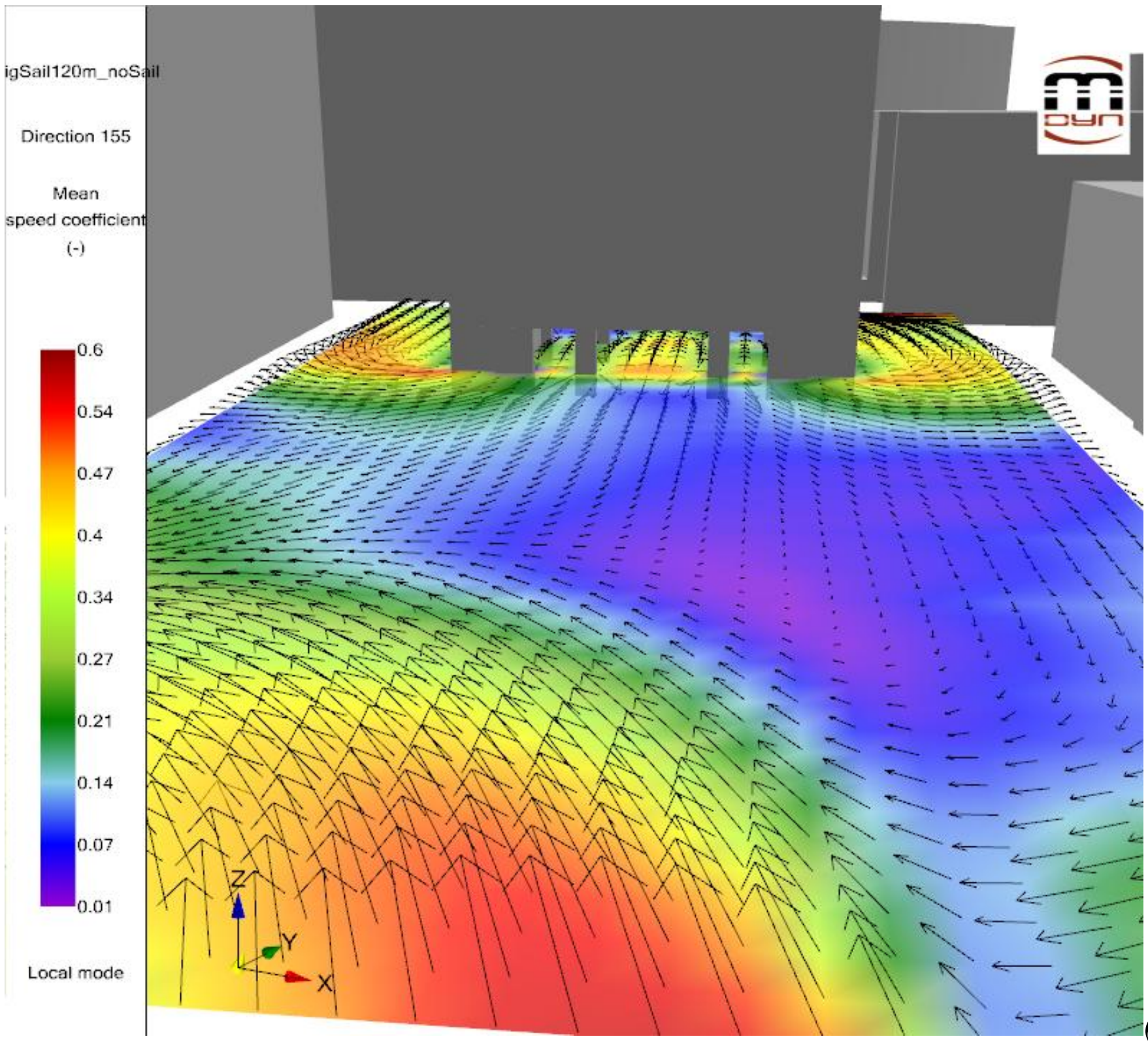

*(c).*

**Figure 4.7.** Wind map at elevation 2 m, perspective view, color shades visualize the wind speed. Cases shown: *(a)* - low porosity permeable stabile, *(b)* - impermeable solid stabile, *(c)* - no stabile.

Figure 4.7 allows a more detailed view of flow around the stabile and a remote view of the Green Building ground level lobby. The top view of this flow is shown in Figure 4.8. The horizontal wind map at 2 m above the ground (Figure 4.8) clearly demonstrates the effect of wind acceleration as it is pushed beneath the base of the Green Building. The southern flow is pushed around the building, generating three separate wind jets - two jets around the sides of the building and a complex central jet through the gap between the building's columns. From Figure 4.8 it is clear that this flow pattern is independent of the details of the local circulation around the stabile.

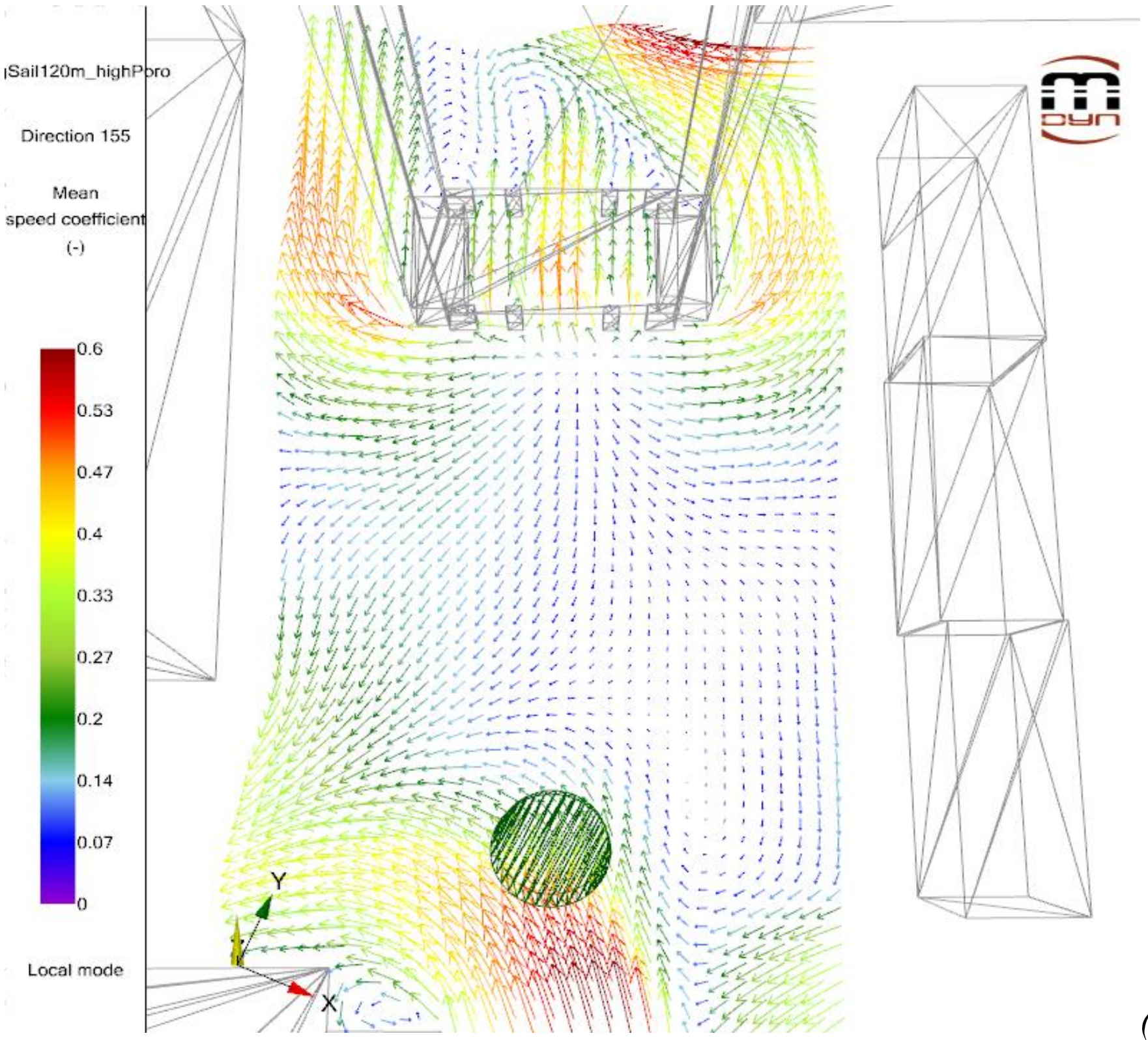

*(a)*.

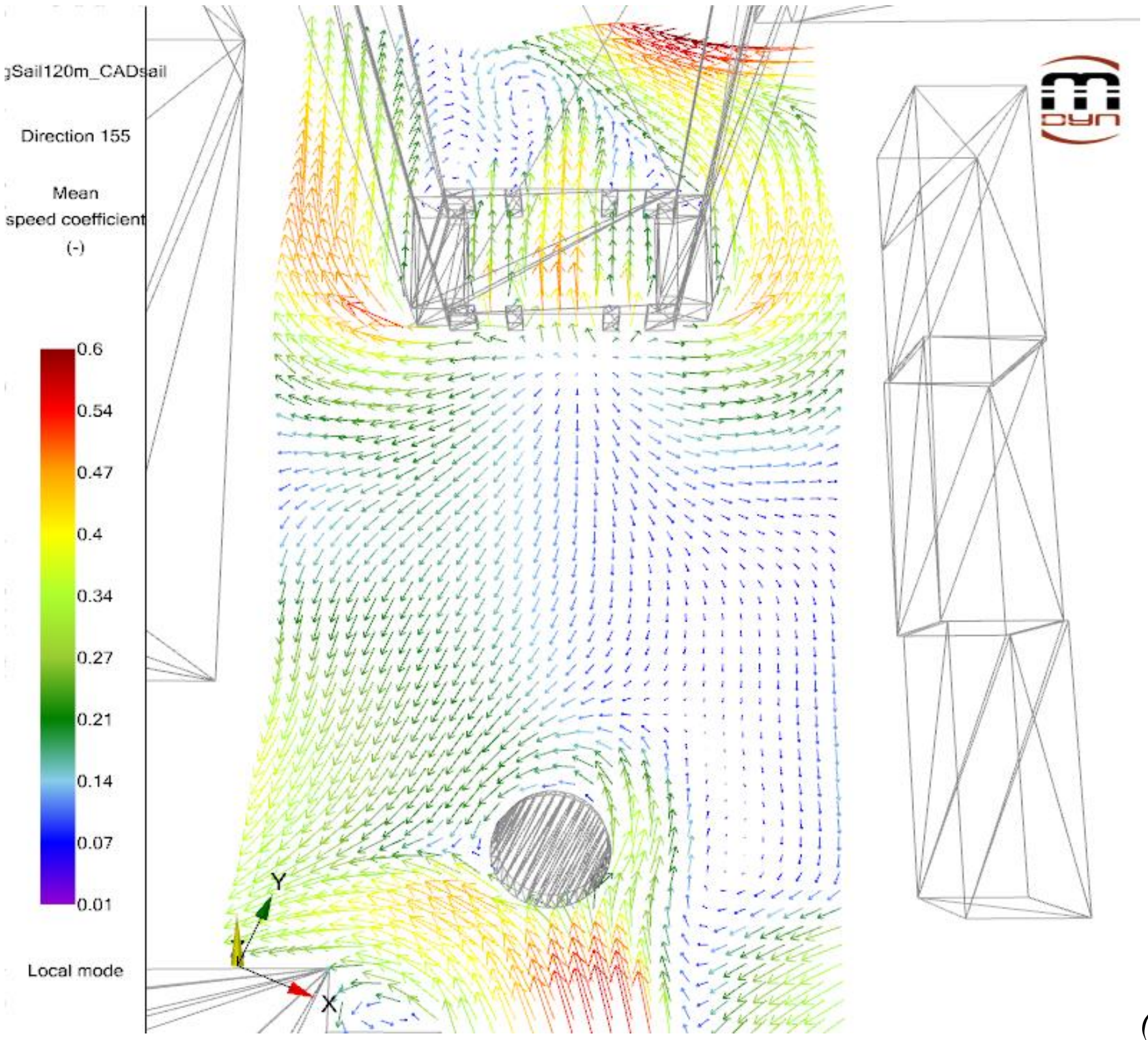

*(b)*.

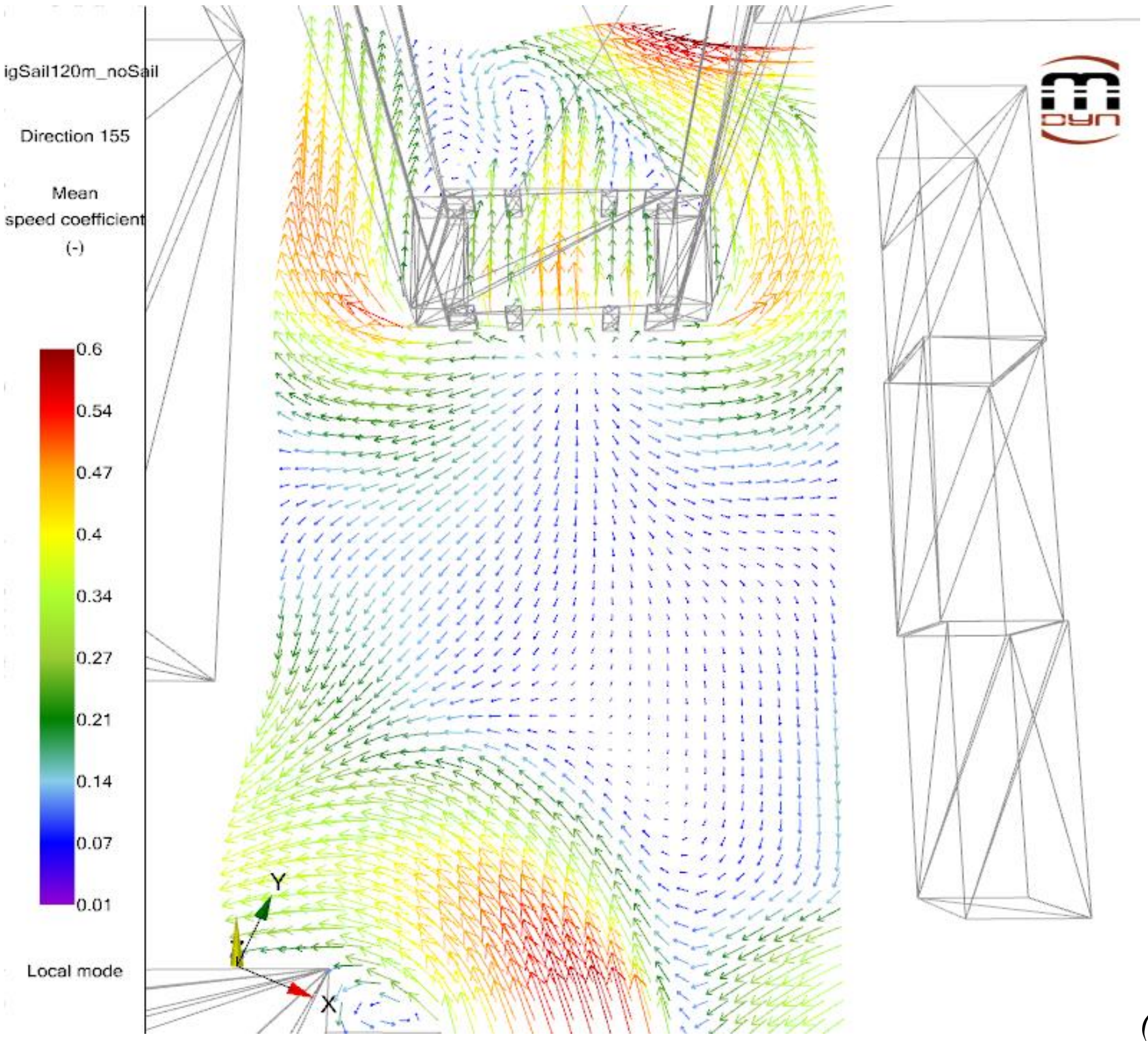

*(c)*.

**Figure 4.8.** Wind vector map at elevation 2 m, top view, vector colors visualize the wind speed. Cases shown: *(a)* - low porosity permeable stabile, *(b)* - impermeable solid stabile, *(c)* - no stabile.

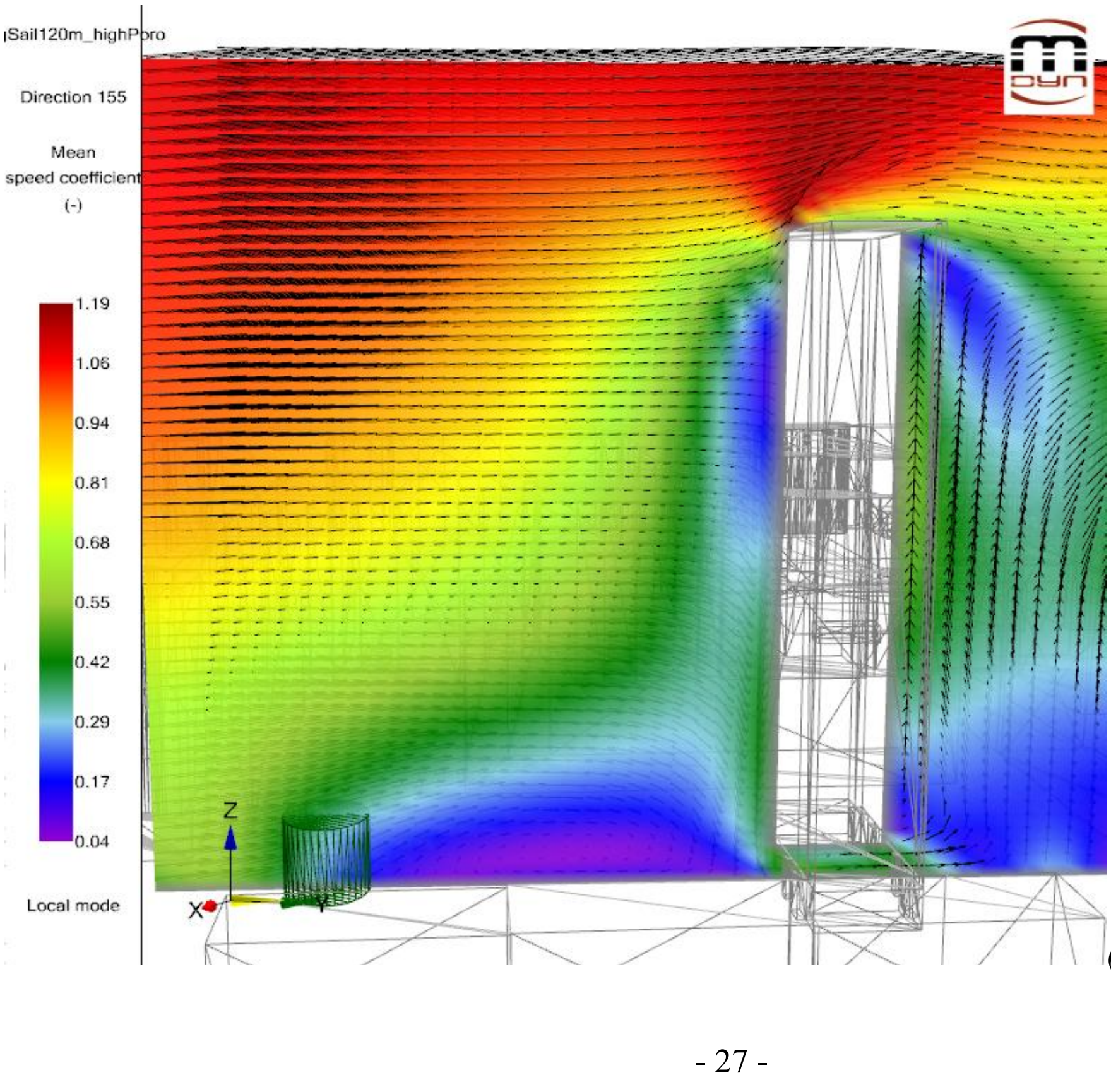

*(a)*.

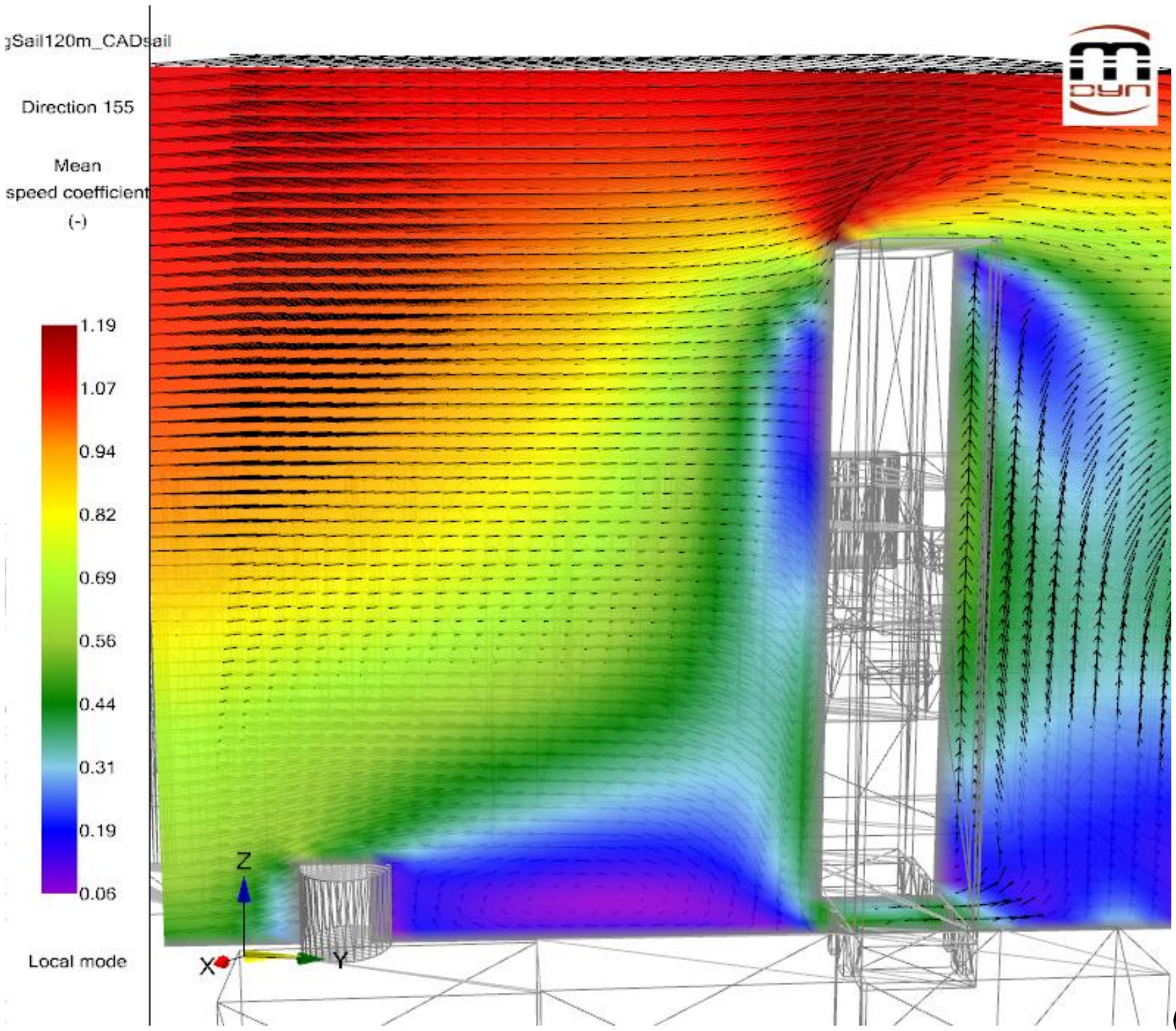

(b).

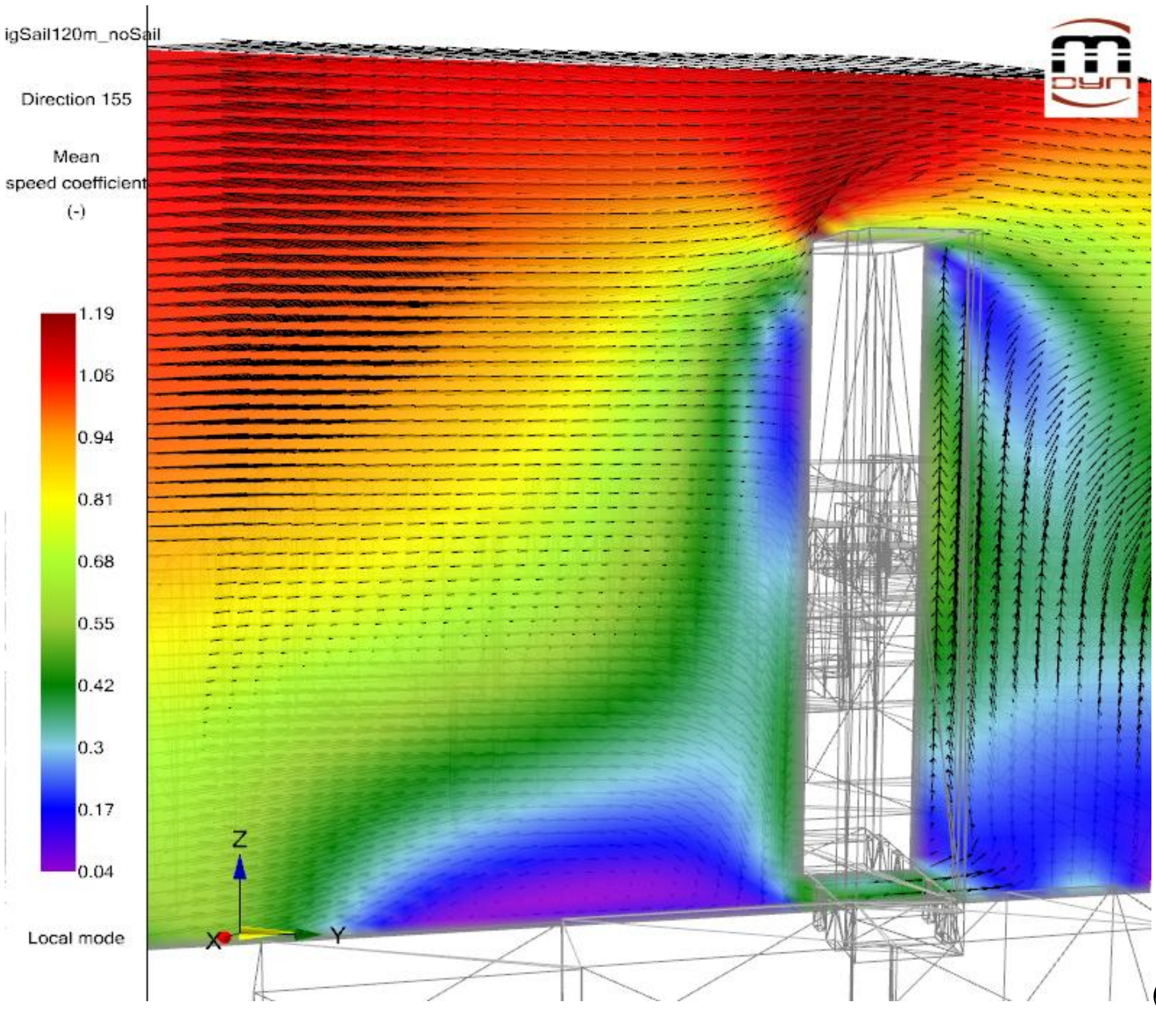

(c).

**Figure 4.9.** Vertical cross section wind map, view from the east, color shades visualize the wind speed. Cases shown: *(a)* - low porosity permeable stabile, *(b)* - impermeable solid stabile, *(c)* - no stabile.

The vertical cross sections in Figure 4.9 and Figure 4.10 reveal the three dimensional structure of the flow. We learn that the weak winds in the open space in front of the Green Building (McDermott Court) are the result of an upwind recirculation zone (see Figure 6 in Shah and Ferziger (1997) for experimental and computational confirmation of such zone existence). The weak low level return flow from the stagnation region in front of the building is capped by a wide horizontally oriented vortex. The vortex extends 10 m above the ground (see also Figure 4.12) and occupies the space between the front of the Green Building and the stabile. It is interesting to observe that the modification of the stabile properties and even its complete removal do not affect the overall pattern. Despite the differences in the structure of the wake behind the stabile, the shape of the recirculation bubble remains very similar. This can be explained by analyzing the pressure distribution. The map of the average pressure perturbation (Figure 4.10) highlights again the minor effect of the stabile on the upwind pressure distribution, while revealing *the key governing mechanism*: The flow around the Green Building is controlled by the large stagnation pressure perturbation at its front face and the corresponding negative perturbation behind it. The magnitude of the positive perturbation reaches more than 100 Pa (1 mbar), while in the flow separation region over the roof (Shah and Ferziger 1997), the pressure drops 160 Pa below the ambient pressure. This pressure structure pushes the air around the building's sides and is responsible for the return flow in the upwind recirculation bubble. It also pushes the air down and creates the jet though the rectangular gap at the base of Green Building. The stagnation pressure at the front face of the building also accelerates the flow over the front edge of the roof (Figure 4.11), qualitatively demonstrating the independence of this effect on flow details around the stabile.

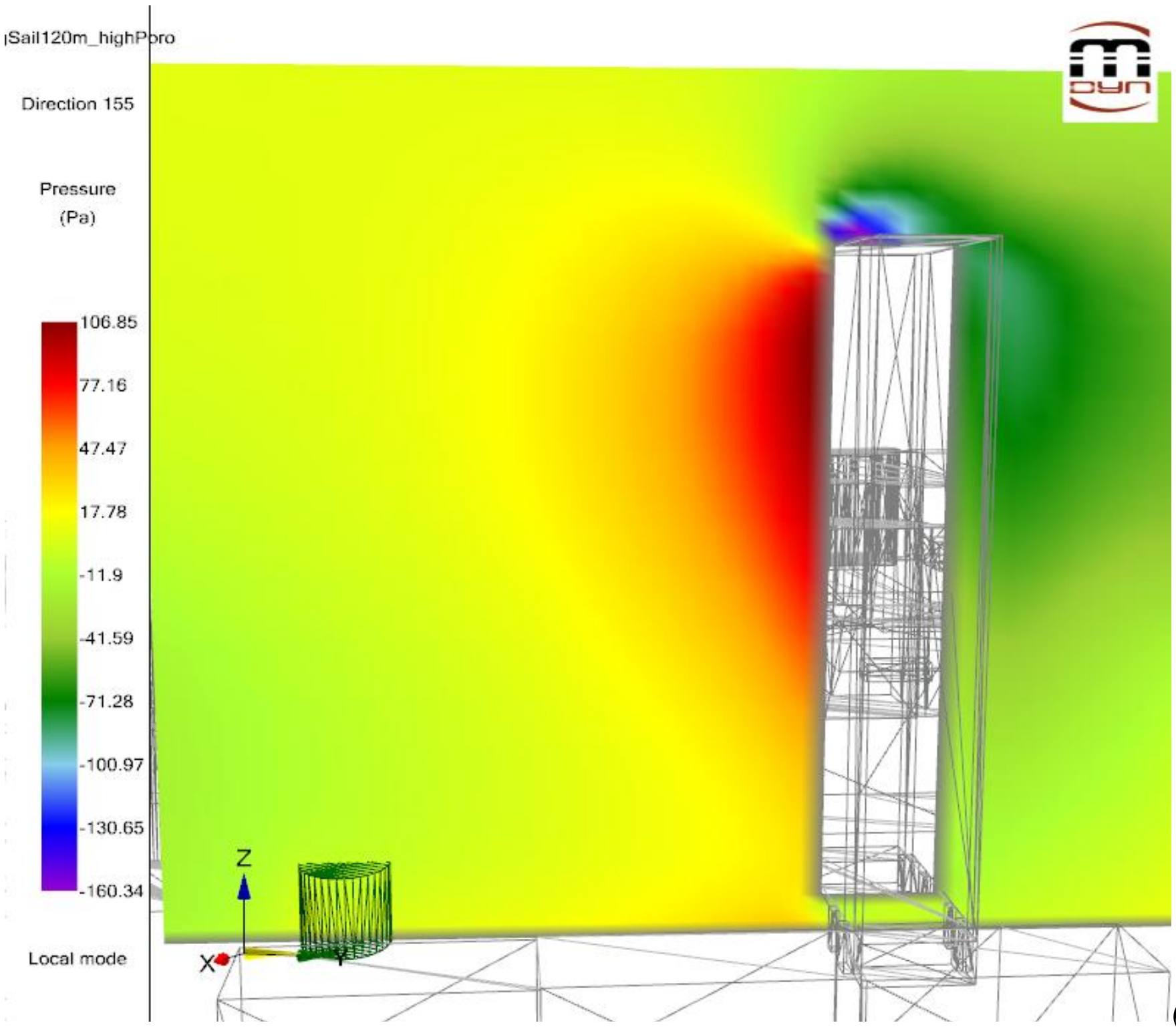
(a).

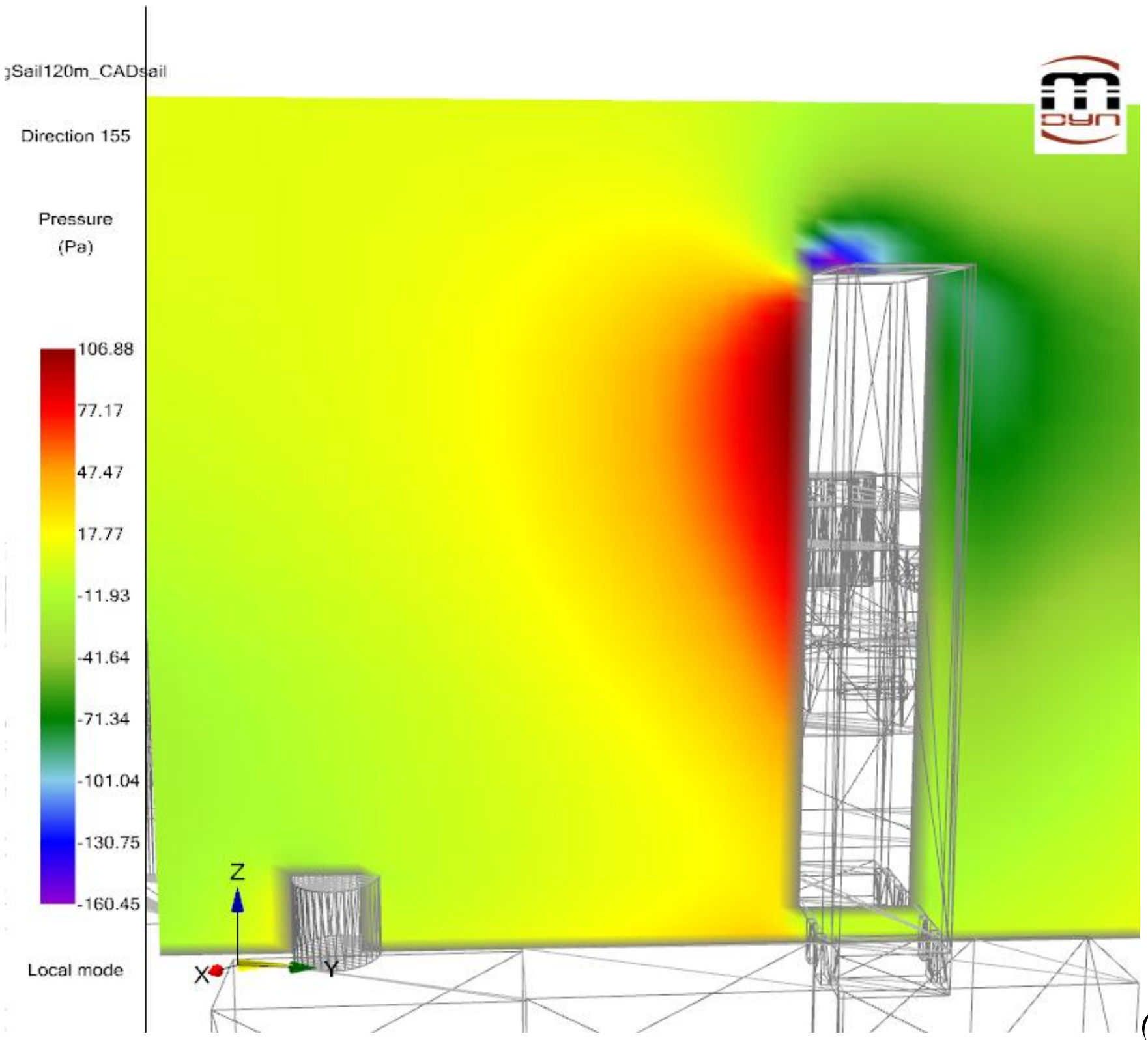
(b).

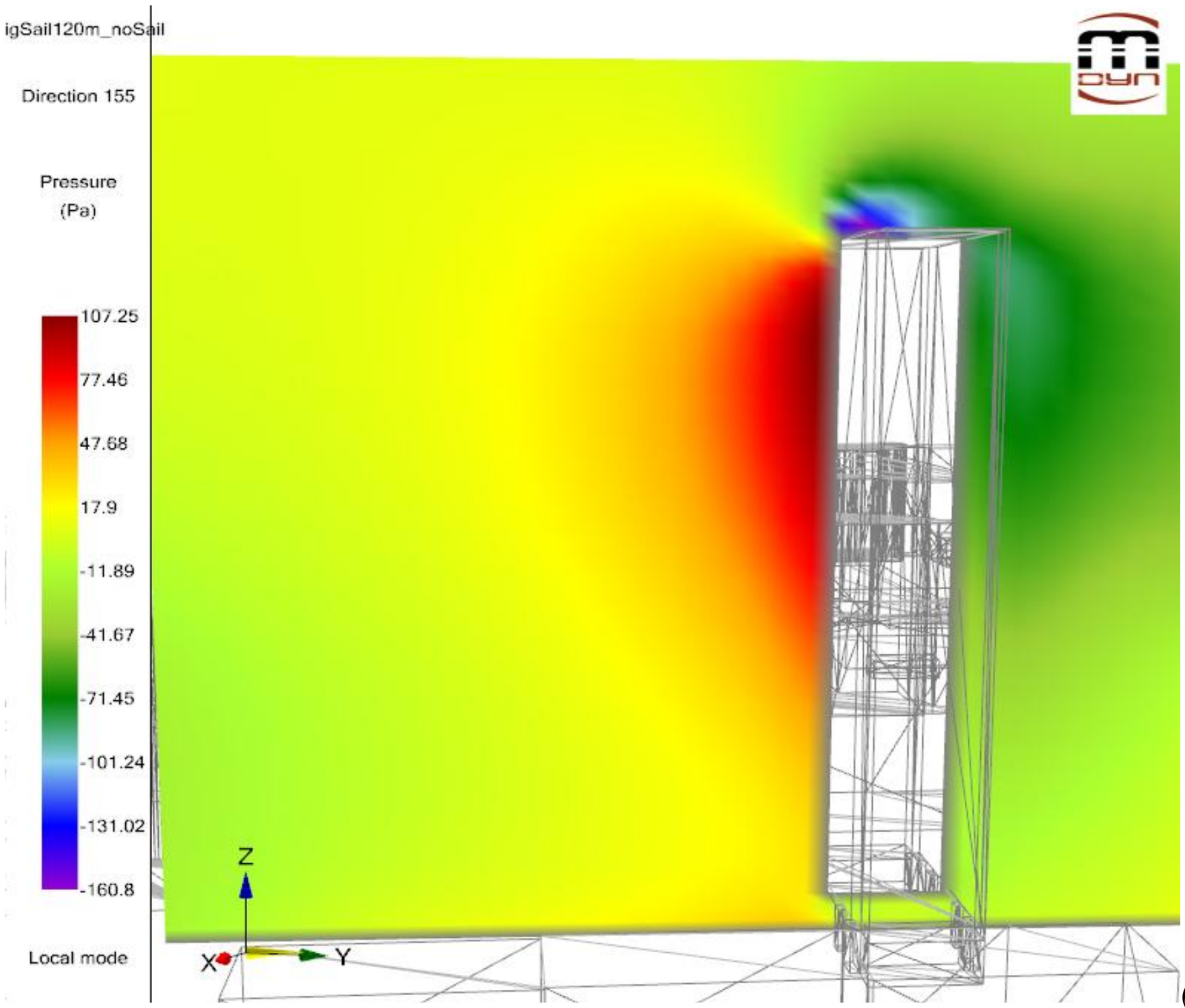

*(c)*.

**Figure 4.10.** Vertical cross section pressure map, view from the east, color shades visualize average pressure perturbation. Cases shown: *(a)* - low porosity permeable stabile, *(b)* - impermeable solid stabile, *(c)* - no stabile.

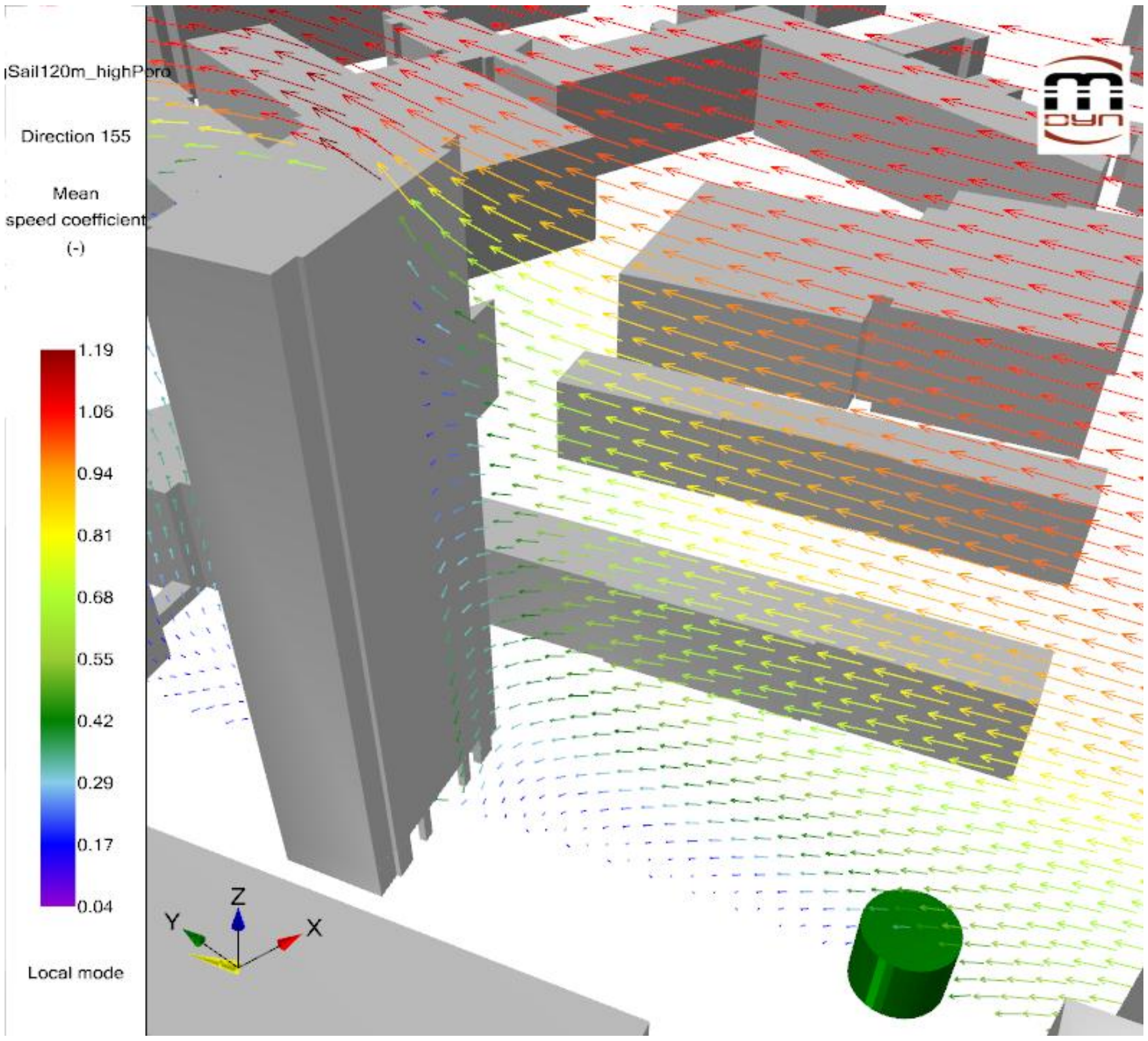

**Figure 4.11.** Vertical cross section wind vector map, view from the west, vector colors visualize the wind speed. Case shown: low porosity permeable stabile.

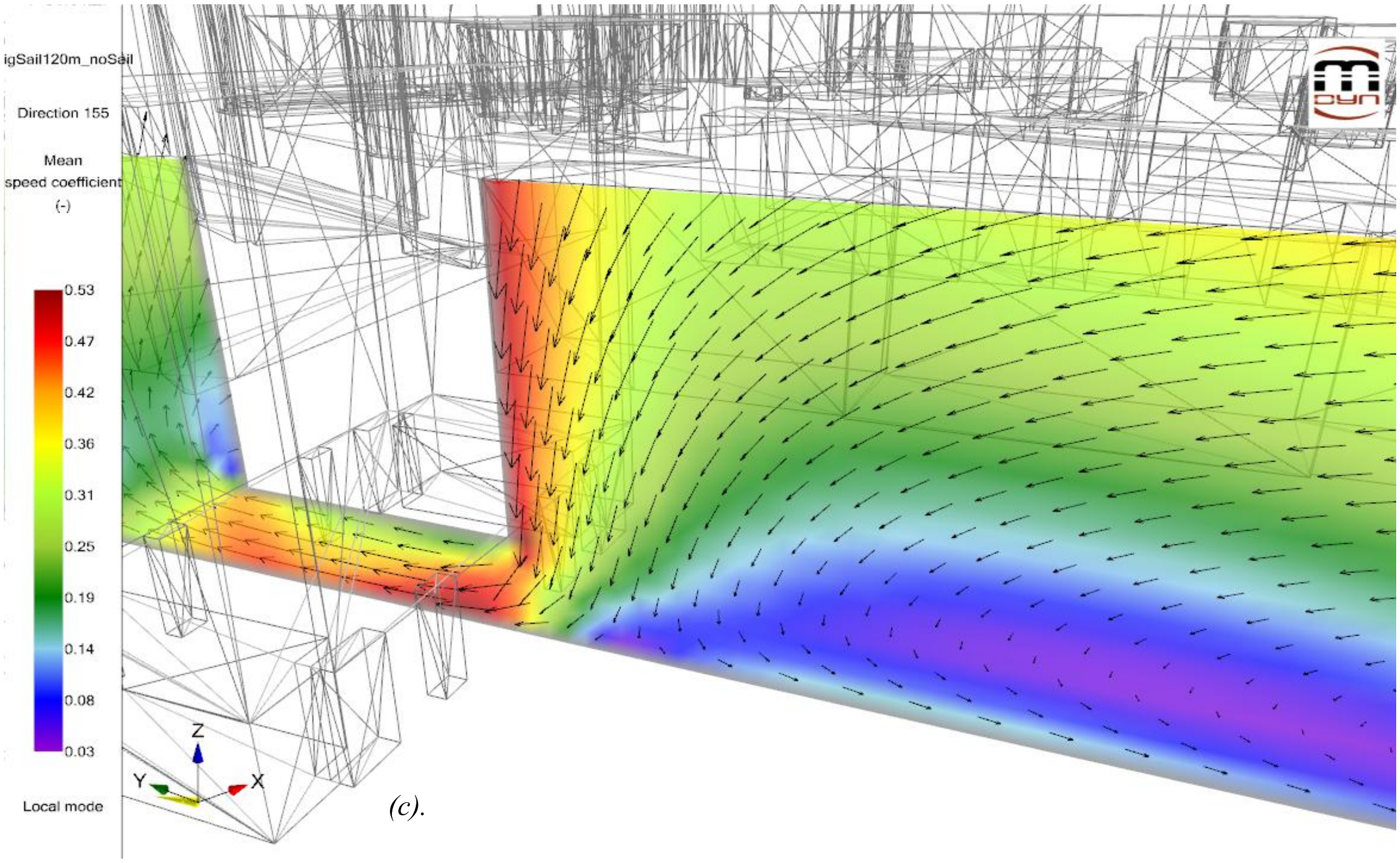

**Figure 4.12.** Vertical cross section wind map, detailed view from the west, color shades visualize wind speed factor. Case shown: no stabile.

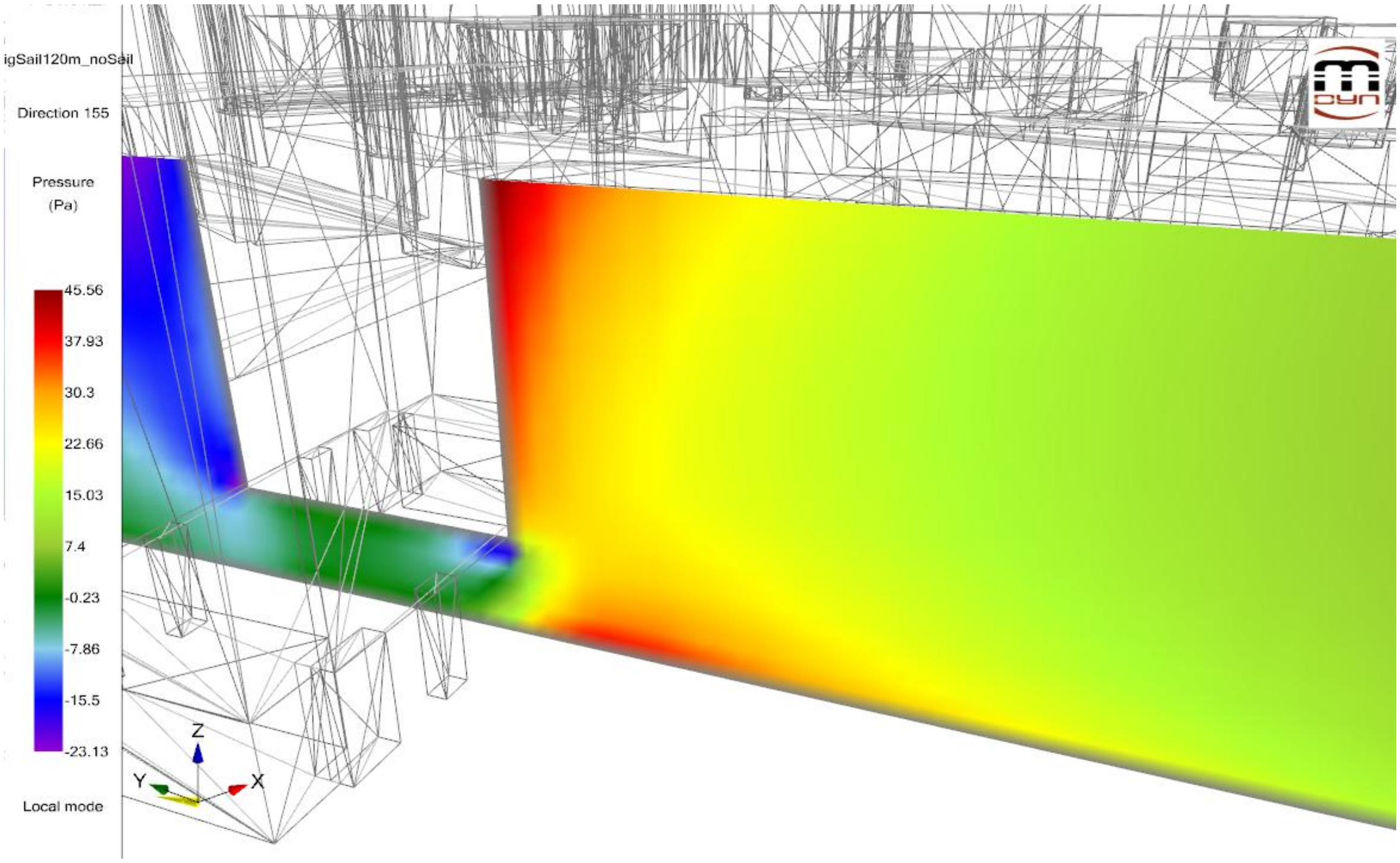

**Figure 4.13.** Vertical cross section pressure map, color shades visualize average pressure perturbation. Case shown: no stabile.

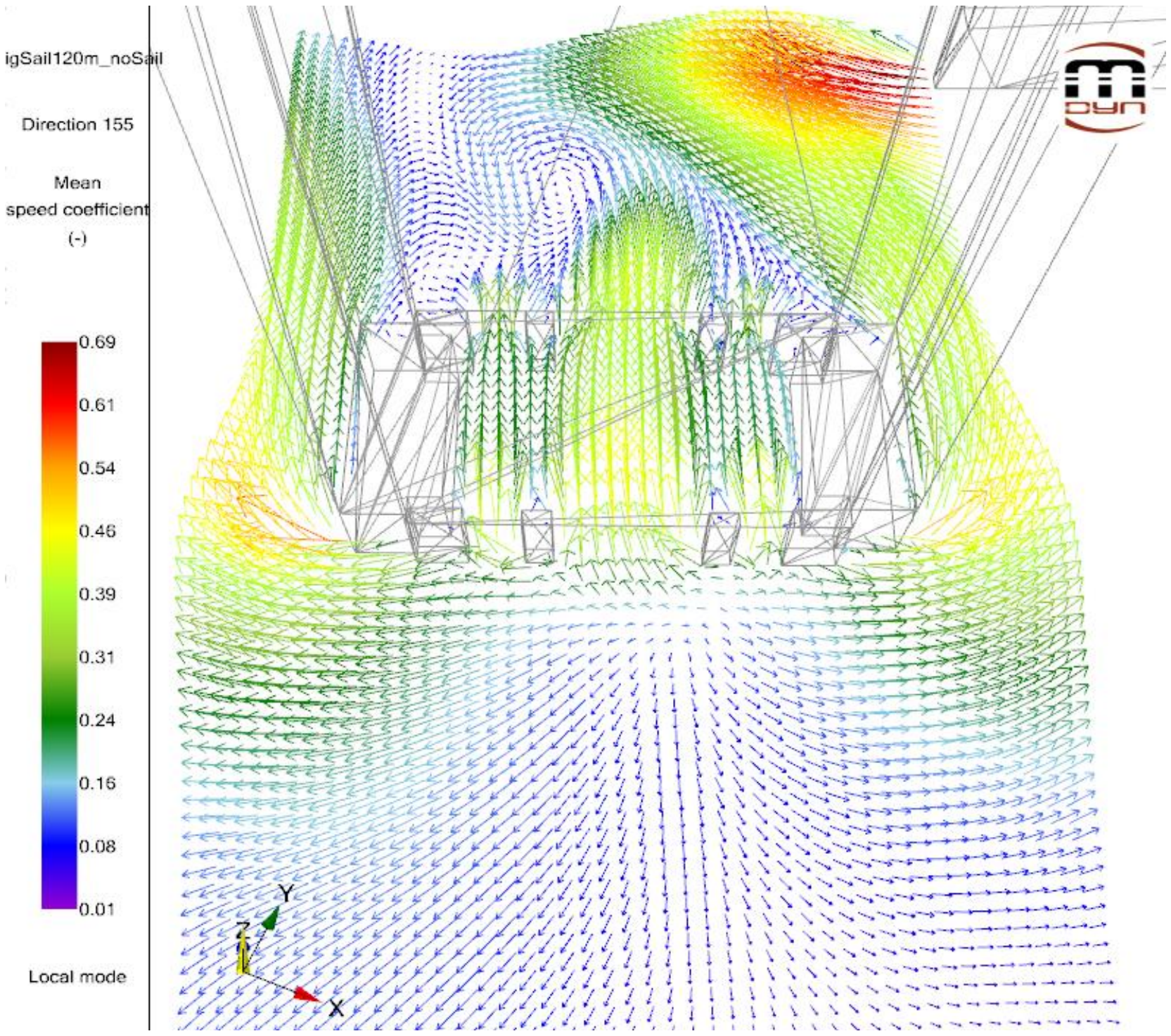

**Figure 4.14.** Detailed view of Green Building lobby jet. Wind vector map at elevation 3 m, vector colors visualize the wind speed. Case shown: no stabile.

Figure 4.12 allows a more detailed view of the vertical structure of the flow through the gap at the base of Green Building. Figure 4.13 highlights the distribution of the pressure in this cross section. It is seen that air is pushed down by the pressure gradient force at the lower part of the front face of the building. After hitting the ground, the flow accelerates in the horizontal direction though the gap and exits the area beneath the building in the form of a jet. Further details of structure of this jet can be seen in Figure 4.14 on the horizontal cross section map, 3 m above the ground.

These CFD results explain the experimental findings in the wind tunnel (Bicknell 1965, 1966) that maximum acceleration of wind flow occurs under the building and near the ground behind it, while the upwind side of Green Building experiences much lower winds than downsteam. Both the hot-wire anemometer measurements (Figure 4.2) and the deflection type indicators (Figure 4.4) illustrate this effect. The weak horizontal upstream winds are due to the stagnation pressure mechanism. Very close to the building the flow is vertical and is not felt near the ground. On the downwind side as well as under the building, the jet accelerates the background flow, causing the pedestrian discomfort observed by the MIT community.

# 5 Summary

Three-dimensional wind flow patterns on the MIT campus were modeled with a CFD model. Wind observations from the roof of the Green Building provided background climatology and were assimilated with the numerical model results. The presented analysis allows a physical insight into flow mechanisms and explains the popular MIT wind myths.

Computed wind fields over the western part of the campus resolve the structure of the flow along the Dorm Row. The presented analysis confirms the common experience of high winds near the corner of MacGregor House. It explains the physical mechanism of this phenomenon as a unique alignment of the prevailing background winds with a long unobstructed acceleration path across the sports field. The stagnation pressure associated with the deceleration of the impinging flow on the northern side of the buildings, further accelerates the wind around the corner of MacGregor House. We demonstrate that for occasional background wind conditions the flow through the Fowler St. channel resembles the structure of a Venturi jet as implied by the name “MacGregor Wind Tunnel.” However, a combined spatial climatological analysis, based on background wind observations, proves that this is not a typical flow regime on average.

We resolve the controversy associated with the effect of "The Big Sail" stabile on the high winds at the doors of the Green Building. It was demonstrated that the strong winds at the base of the Green Building are the result of a large stagnation pressure perturbation at its front face. CFD simulations show the detailed structure of the wind jet in the gap at the base of the Green Building and the recirculation vortex over the McDermott Court. We have proved that the existence of the "The Big Sail" stabile does not modify significantly the patterns of the flow around the Green Building and confirmed that as the MIT arts administrator Bill Arning says (Gleitzman 2006): "The wind effect (on the Green Building doors) was not altered in any way by the location of the `Big Sail`".

# 6 Acknowledgments

The CFD study of wind on MIT campus was supported by the MIT Energy Initiative Student Grant. I thank the staff of MIT GIS Lab and MIT Synoptic Laboratory for the technical support in accessing the GIS and meteorological data. I am grateful to professor Leslie K. Norford for his advice and to Pamela Siska for helping me edit this manuscript. Special thanks to the other members of WEPA and Project Full Breeze for the exciting opportunity to engage in Wind Energy projects on MIT campus.